\documentclass[twocolumn]{autart}

\usepackage{tikz}

\usepackage{amsmath}
\usepackage{latexsym, amssymb}
\usepackage{graphicx}
\usepackage{amsmath, amsbsy}
\usepackage{amsopn, amstext}
\usepackage{hyperref}
\usepackage{cancel, color}
\usepackage{epstopdf}

\def\J{{\bf 1}}

\DeclareMathOperator{\Col}{Col}
\DeclareMathOperator{\Row}{Row}

\DeclareMathOperator{\tr}{Trace}

\DeclareMathOperator{\lcm}{lcm}

\def\cal{\mathcal}

\def\diag{diag}
\def\PR{PR}

\def\diag{diag}
\def\ra{\rightarrow}

\def\a{\alpha}
\def\b{\beta}
\def\d{\delta}

\def\D{\Delta}

\def\0{{\bf 0}}
\def\J{{\bf 1}}

\newcommand{\R}{{\mathbb R}}

\newcommand{\Z}{{\mathbb Z}}

\def\dsum{\mathop{\sum}\limits}
\def\thefootnote{*}

\newtheorem{dfn}[thm]{Definition}
\newtheorem{prp}[thm]{Proposition}
\newtheorem{exa}[thm]{Example}

\begin{document}

\begin{frontmatter}

\title{On networks over finite rings\thanksref{footnoteinfo}}

\thanks[footnoteinfo]{This paper was not presented at any IFAC
meeting. This work is supported partly by NNSF
  62073315, 61074114, and 61273013 of China. Corresponding author: Daizhan Cheng. Tel.: +86 10 6265 1445; fax.: +86 10 6258 7343.}

\author[AMSS,STPC]{Daizhan Cheng}\ead{dcheng@iss.ac.cn}

\address[AMSS]{Key Laboratory of Systems and Control, Academy of Mathematics and Systems Science, Chinese Academy of Sciences, Beijing 100190, P.R.China}
\address[STPC]{Research Center of Semi-tensor Product of Matrices, Theory and Applications, Liaocheng University, Lianocheng, P.R. China}


\begin{abstract}
A (control) network over a finite ring is proposed. Using semi-tensor product (STP) of matrices, a set of algebraic equations are provided to verify whether a finite set with two binary operators is a ring. It is then shown that the STP-based technique developed for logical (control) networks are applicable to (control) networks over finite rings. The sub- (control) network over an ideal of the bearing ring is revealed. Then the product ring is proposed and the (control) network over product ring is investigated. As a key result, the decomposition theorem, called the Decomposition Principle (DP), is proposed, which shows a (control) network over a product ring is decomposable into sub- (control)networks over each factor rings, which makes the (control) properties of a network can be revealed by its factor sub-networks over factor rings. Using DP, the control problems of a control network are investigated via its sub control networks. Particularly, using DP, the control of linear networks over product rings  is discussed in detail. Finally, the representation theorem is presented, which shows that each (control) network over a finite set can be expressed as a (control) network over a product ring.
\end{abstract}

\end{frontmatter}

\section{Introduction}

In recent years, the networks over finite sets attract a considerable attention from control community \cite{kas07,yoo21}. As pointed by \cite{men20}, ``finite sets are more realistic in dealing with limited storage capacities and communication bandwidths". Among these finite set approaches, networks over finite fields are of particular interest \cite{sun12,pas14,li16,li18,li19,men20}.

Finite field, also called Galois field, is of the form $\Z_p$ where $p$ is a primer number \cite{lan02}.
From existing results one sees easily that finite field approach has several advantages, such as reducing the computational complexity \cite{men20}, using semi-tensor product (STP) technique \cite{li19}, easily applicable to other problems \cite{sun12,pas14,li16}. In addition, the field structure provides convenience in theoretical analysis.

An obvious restriction on finite field approach is that the cardinal number (number of elements) of a finite set considered should be a prime number. This restriction is practically unreasonable. By the author's understanding, it comes from technical point of view. That is, the treatment tool for finite fields is ready and convenient.  A natural generalization is: instead of finite fields can we consider certain more general structures, say, $\Z_r$, for any integer $r\geq 2$ to replace finite fields? Hence, this paper considers the networks over finite rings which contain finite fields or even  $\Z_r$ with any $r>3$ as  special cases. This is the original motivation for this paper.

Recently, STP of matrices has been proposed and used to analysis and control design of Boolean networks as well as $k$-valued logical networks \cite{che11,che12}. STP was first used to investigate networks over finite field by \cite{li19}. STP is also a fundamental tool for the investigation in this paper.

After a brief survey on rings and the STP approach to logical networkers, this paper provides a set of algebraic equations for finite rings, which can be used to construct finite rings and investigate their properties. After a finite and exhausting searching, we found that there are exactly $6$ rings of cardinal number $4$.  Then we show that the STP-based techniques developed for logical (control) networks are applicable to (control) networks over finite rings. When $S$ is an ideal of the bearing ring $R$ of a network, it is shown that a sub-network is obtained over $S$, which can be considered as a sub-network over an invariant subspace. Motivated by this, the product ring of some fundamental rings is proposed. All the factor rings are shown to be ideals of the product ring. A general product ring $\Z^{\kappa}$ is proposed. When $\kappa=\prod_{i=1}^sk_i$ is a prime number decomposition, $\Z^{\kappa}$ is the product ring of $\Z_{k_i}$, $i\in [1,s]$. Using product ring, a key result, called the decomposition principle (DP), is presented, which says each (control) network over $\Z^{\kappa}$ can be decomposed into $s$ sub- (control) networks over its component sub-rings $\Z_{k_i}$, which are finite fields. Decomposition principle allows us to understand and design control for original networks by analyzing and controlling all component sub-networks. In addition, it also makes all  results for finite fields applicable to finite rings. As an application, the control of networks over finite rings is considered by designing controls for all factor sub networks. Particularly, using DP, the linear control networks over $\Z^{\kappa}$ are discussed in detail. Finally, another key result, called the representation theorem, is presented, which shows that any (control) network over finite sets, say $\Xi$, can be expressed as a (control) network over $\Z^{\kappa}$, where $\kappa=|\Xi|$. That is, the technique developed for networks over finite rings is universally applicable to any networks over finite sets, including Boolean (control) networks, $k$-valued or mix-valued logical (control) networks, (control) networks over finite fields, shift registers \cite{zho16}, and strategy profile dynamics of finite games, etc.

The rest of this paper is organized as follows: Section 2 is a brief survey for finite rings and STP technique to logical (control) networks. Using STP, Section 3 provides a set of algebraic equations for finite rings. As an application, all the finite rings with $4$ elements are obtained. Section 4 shows that the STP-based technique developed for logical (control) networks are applicable to networks over finite rings. For a given (control) network over a ring the sub-network over an ideal is investigated in Section 5. Section 6 proposes the product ring of finite rings. The networks over product rings are also proposed and studied. A key theorem, called decomposition principle, is presented in Section 7, which shows that a network over a product ring can be decomposed into sub-networks on each factor rings.
Section 8 considers the control of networks over product rings. As a special case of Section 8, the decomposition of linear (control) networks over product rings are investigated  in Section 9. Section 10 presents another key theorem, called representation theorem, which shows any (control) networks over finite sets can be expressed as  networks over  product rings. Section 11 is a brief conclusion.

\section{Preliminaries}

\subsection{Rings and Sub-Rings}

\begin{dfn}\label{d.1.1.1}\cite{lan02} Let $R\neq \emptyset$. $R$ is called a ring if there exist two binary operators $(\oplus,\odot)$ such that
\begin{itemize}
\item[(i)] $(R,\oplus)$ is an Abelian group with ${\bf 0}$ as its identity element.
\item[(ii)] $(R,\odot)$ is a monoid, i.e., a semi-group with identity $\J \neq {\bf 0}$.
\item[(iii)]
\begin{align}\label{1.1.1}
\begin{array}{l}
(a\oplus b)\odot c=a\odot c\oplus b\odot c,\\
a\odot (b\oplus c)=a\odot b\oplus a\odot c,\quad a,b,c\in R.
\end{array}
\end{align}
\end{itemize}
In addition, if
\begin{align}\label{1.1.2}
 a\odot b=b\odot a,\quad a,b\in R,
\end{align}
then $R$ is called  a commutative ring.
\end{dfn}

\begin{rem}\label{r.1.1.2}~~
\begin{itemize}

\item[(i)] In some literature a ring may have no (product) identity. Here we follow the definition in \cite{lan02}, and consider  a ring with identity as a ring. To emphasize the operators on $R$ we sometimes also express $R$ as a triple $(R,\oplus,\odot)$.
\item[(ii)] If $R$ is a ring and $(R\backslash \{0\},\times)$ is also an Abelian group, then $R$ is a field.
Hence a finite field is a special finite ring.
\item[(iii)] Since $(R,\oplus)$ is a group, for each $X\in R$ there exists its unique inverse, denoted by $\neg X$ (or $-X$) such that $X\oplus (\neg X)={\bf 0}$. We also use $\ominus$ as $X\ominus Y:=X\oplus (\neg Y)$.
\end{itemize}
\end{rem}

Some fundamental properties of a ring are as follows:

\begin{prp}\label{p.1.1.3} \cite{hun74} Let $R$ be a ring with $\J\neq {\bf 0}$. Then
\begin{itemize}
\item[(i)]
\begin{align}\label{1.1.3}
 {\bf 0}\odot a=a\odot {\bf 0}={\bf 0},\quad a\in R.
\end{align}
\item[(ii)]
\begin{align}\label{1.1.4}
(\neg a)\odot b=a\odot (\neg b)=\neg (a\odot b),\quad a,b\in R.
\end{align}
\item[(iii)]
\begin{align}\label{1.1.5}
(na)\odot b=a\odot (nb)=n(a\odot b),\quad n\in \Z_+, \;a,b\in R.
\end{align}
Note that $na:=\underbrace{a\oplus a\oplus \cdots\oplus a}_n$.
\item[(iv)]
\begin{align}\label{1.1.6}
\left(\oplus_{i=1}^ma_i\right) \left(\oplus_{j=1}^nb_j\right)=\oplus_{i=1}^m\oplus_{j=1}^na_i\odot b_j,\quad a_i,b_j\in R.
\end{align}
\end{itemize}
\end{prp}

\begin{dfn}\label{d.1.1.4} Let $(S,+,\times)$ and $(R,\oplus,\otimes)$ be two rings.
\begin{itemize}
\item[(i)] A mapping $\pi:S\ra R$ is called a ring homomorphism, if
\begin{align}\label{1.1.3}
\begin{cases}
\pi(s_1+s_2)=\pi(s_1)\oplus \pi(s_2)\\
\pi(s_1\times s_2)=\pi(s_1)\otimes \pi(s_2),\quad s_1,s_2\in S.\\
\end{cases}
\end{align}
If there exists a homomorphism $\pi:S\ra R$, then $S$ is said to be homomorphic to $R$, denoted by $S\simeq R$.

\item[(ii)] Let $\pi:S\ra R$ be a ring homomorphism. $\pi$ is called a ring isomorphism if $\pi$ is a bijective mapping. Then $S$ and $R$ are said to be isomorphic, denoted by $S\cong R$.

\item[(iii)] If $\pi:R\ra R$ is a ring isomorphism, it is called an automorphism.
\end{itemize}
\end{dfn}

\begin{rem}\label{r.1.1.4} Let $S$ and $R$ be two rings.

\begin{itemize}
\item[(i)]
 If $\pi:S\ra R$ is a ring homomorphism, then it is easy to see that $\pi(1)=1$, $\pi(0)=0$, and
$$
\pi[(s_1+s_2)\times s_3]=\pi(s_1)\odot \pi(s_3)\oplus \pi(s_2)\odot \pi(s_3),
$$
etc.
\item[(ii)] For inverse element $\neg x$, we have
$$
\pi(\neg x)=\neg \pi(x).
$$
\end{itemize}
\end{rem}

\begin{dfn}\label{d.1.1.5}~~
\begin{itemize}
\item[(i)] Let  $(R,\oplus,\odot)$ be a ring. $S\subset R$. If $(S,\oplus,\odot)$ is also a ring, $S$ is called a sub-ring of $R$.

\item[(ii)] A sub-ring $S\subset R$ is said to be an ideal of $R$, if
$$
r\odot S\subset S,\quad S\odot r\subset S,\quad r\in R.
$$
$r\odot S$ ($S\odot r$) is called the left (right) coset of $S$. If $R$ is commutative, then $r\odot S=S\odot r$.
\end{itemize}
\end{dfn}

\begin{exa}\label{e.1.1.6}

Consider $\Z_p$. The operators
$$
a +_p b:=a+b (\mbox{mod}~p),\quad a \times_p b:=a\times b (\mbox{mod}~p).
$$

\begin{itemize}
\item[(i)] Consider $\Z_6$. Let $S=\{3,0\}$.
Then it is easy to verify that $(S,+_6,\times_6)$ is a ring. Hence, $S$ is a sub-ring of $\Z_6$.

\item[(ii)] Let $\pi:S\ra \Z_2$, defined by
$$
\pi(x)=
\begin{cases}
1,\quad x=3\\
0,\quad x=0.
\end{cases}
$$
Then it is easy to see that $\pi:S\ra \Z_2$ is an isomorphism.

\item[(iii)] It is clear that $S$ is an ideal of $\Z_6$.
\end{itemize}

\end{exa}

Let $|S|=r$. $S$ is isomorphic to a ring $(E,\oplus_E,\odot_E)$, where $E=\{1,2,\cdots,r-1.0\}$, then $E$ is called the essential ring of $S$.

Consider $S=\{3,0\}\subset \Z_6$. The essential ring of $S$ is $\Z_2$.

\subsection{STP Approach to Logical Networks}

The following is a brief survey for STP. We refer to \cite{che11,che12} for more details.

\begin{dfn}\label{d.1.2.1} Let $M\in {\cal M}_{m\times n}$ and $N\in {\cal M}_{p\times q}$, $t=\lcm(n,p)$ be the least common multiple of $n$ and $p$. Then the STP of $M$ and $N$ is defined by
\begin{align}\label{1.2.1}
A\ltimes B:=\left(A\otimes I_{t/n}\right)\left(B\otimes I_{t/p}\right),
\end{align}
where $\otimes$ is the Kronecker product.
\end{dfn}

\begin{rem}\label{r.1.2.2}~~~
\begin{itemize}
\item[(i)] If $n=p$, then $A\ltimes B=AB$. Hence the STP is a generalization of the conventional matrix product.
\item[(ii)] STP keeps all the main properties of the conventional matrix product available. Hence in most cases the symbol $\ltimes$ is omitted if there is no possible confusion.
\item[(iii)] Hereafter, all the matrix products in this paper are assumed to be SPT unless elsewhere stated.
\end{itemize}
\end{rem}

The following property is useful in the sequel.

\begin{prp}\label{p.1.2.3} Let $M$ be a matrix, $X\in \R^t$ be a column vector. Then
\begin{align}\label{1.2.2}
XM=(I_t\otimes M)X.
\end{align}
\end{prp}

The swap matrix $W_{[m,n]}$ is defined as
\begin{align}\label{1.2.3}
W_{[m,n]}=[I_n\d_m^1,I_n\d_m^2,\cdots,I_n\d_m^m],
\end{align}
where $\d_m^i$ is the $i$ th column of identity matrix $I_m$.

The main function of a swap matrix is shown in the following proposition.

\begin{prp}\label{p.1.2.4} Let $X\in \R^m$ and $Y\in \R^n$ be two column vectors.  Then
\begin{align}\label{1.2.4}
W_{[m,n]}XY=YX.
\end{align}
\end{prp}

Next, we briefly survey the STP based technique for logical (control) networks.

Let ${\cal D}_k=\{1,2,\cdots,k-1,0\}$ be a finite set. It can be used to represent any finite set $S$ with $|S|=k$. Denote by $\D_k:=\Col(I_k)$ the set of columns of the identity matrix $I_k$, then the elements in ${\cal D}_k$ can be expressed by vectors as
\begin{align}\label{1.2.401}
\vec{i}=
\begin{cases}
\d_k^i,\quad 1\leq i\leq k-1\\
\d_k^k,\quad i=0.
\end{cases}
\end{align}

A matrix $L\in {\cal M}_{m\times n}$ is called a logical matrix, if its columns
$\Col(L)\subset \D_m$. Then it can be expressed as
$$
L=\d_m[i_1,i_2,\cdots,i_n]:=[\d_m^{i_1},\d_m^{i_2},\cdots,\d_m^{i_n}].
$$
The set of $m\times n$ logical matrices is denoted by ${\cal L}_{m\times n}$.

\begin{prp}\label{p.1.2.5} Consider a  $k$-valued logical function
\begin{align}\label{1.2.5}
Y=f(X_1,X_2,\cdots,X_n),
\end{align}
where $X_i\in {\cal D}_k$, $i\in [1,n]$. Then there exists a unique $M_f\in {\cal L}_{k\times \kappa}$, ($\kappa=k^n$), called
the structure matrix of $f$, such that
\begin{align}\label{1.2.6}
y=M_f\ltimes_{i=1}^nx_i,
\end{align}
where $y=\vec{Y}$ and $x_i=\vec{X}_i$, $i\in [1,n]$, are vector forms of $Y$ and $X_i$ respectively.
\end{prp}

Using Proposition \ref{p.1.2.5}, a logical dynamic (control) network can be expressed into its vector form as follows:

\begin{prp}\label{p.1.2.6}
Consider a $k$-valued logical network
\begin{align}\label{1.2.7}
\begin{cases}
X_1(t+1)=f_1(X_1(t),X_2(t),\cdots,X_n(t))\\
X_2(t+1)=f_2(X_1(t),X_2(t),\cdots,X_n(t))\\
\vdots\\
X_n(t+1)=f_n(X_1(t),X_2(t),\cdots,X_n(t)).\\
\end{cases}
\end{align}
Then in vector form it can be expressed as
\begin{align}\label{1.2.8}
\begin{cases}
x_1(t+1)=M_1x(t)\\
x_2(t+1)=M_2x(t)\\
\vdots\\
x_n(t+1)=M_nx(t),\\
\end{cases}
\end{align}
where $x(t)=\ltimes_{i=1}^nx_i(t)$. $M_i$ is the structure matrix of $f_i$. (\ref{1.2.8}) is called the component-wise algebraic state space representation (ASSR) of (\ref{1.2.7}).

Moreover, (\ref{1.2.8}) can further be expressed as
\begin{align}\label{1.2.9}
x(t+1)=Mx(t),
\end{align}
where
$$
M=M_1*M_2*\cdots*M_n,
$$
where $*$ is the  Khatri-Rao product \cite{che12}.

(\ref{1.2.9}) is called the ASSR of (\ref{1.2.7}).
\end{prp}

\begin{prp}\label{p.1.2.7}
Consider a $k$-valued logical control network
\begin{align}\label{1.2.10}
\begin{array}{l}
\begin{cases}
X_1(t+1)=f_1(X_1(t),\cdots,X_n(t),U_1(t),\cdots,U_m(t))\\
X_2(t+1)=f_2(X_1(t),\cdots,X_n(t),U_1(t),\cdots,U_m(t))\\
\vdots\\
X_1(t+1)=f_n(X_1(t),\cdots,X_n(t),U_1(t),\cdots,U_m(t)),\\
\end{cases}\\
~~~Y_{\ell}=g_{\ell}(X_1(t),X_2(t),\cdots,X_n(t)),\quad \ell \in[1,p].
\end{array}
\end{align}
Similarly to logical case, its component-wise ASSR is
\begin{align}\label{1.2.11}
\begin{array}{l}
\begin{cases}
x_1(t+1)=L_1u(t)x(t)\\
x_2(t+1)=L_2u(t)x(t)\\
\vdots\\
x_n(t+1)=L_nu(t)x(t),\\
\end{cases}\\
~~~y_{\ell}(t)=E_{\ell}x(t),\quad \ell\in [1,p],
\end{array}
\end{align}
where $u(t)=\ltimes_{j=1}^mu_j(t)$, $L_i$ and $E_{\ell}$ are structure matrices of $L_i$ and $g_{\ell}$ respectively.

Moreover, (\ref{1.2.11}) can further be expressed into its ASSR form as
\begin{align}\label{1.2.12}
\begin{cases}
x(t+1)=Lx(t),\\
y(t)=Ex(t),
\end{cases}
\end{align}
where
$$
\begin{array}{l}
L=L_1*L_2*\cdots*L_n,\\
E=E_1*E_2*\cdots*E_p.
\end{array}
$$
\end{prp}

A useful tool for deducting ASSR of logical (control) networks is the following matrix, called a power reducing matrix:
\begin{align}\label{1.2.13}
\PR_n:=\diag(\d_n^1,\d_n^2,\cdots,\d_n^n)\in {\cal L}_{n^2\times n}.
\end{align}

\begin{prp}\label{p.1.2.8} Let $x\in \D_n$. Then
\begin{align}\label{1.2.14}
\PR_nx^2=x, \quad x\in \D_n.
\end{align}
\end{prp}

\section{Matrix Expression of Operators on Finite Rings}

Consider a finite set $R=\{1,2,\cdots,k-1,0\}$. Assume it has two binary operators $\oplus$ and $\odot$. Using vector form expression (\ref{1.2.401}), $R\equiv \D_k$.
Assume the structure matrices for $\oplus$ and $\otimes$ are $M_{\oplus},~M_{\odot}\in {\cal L}_{k\times k^2}$ respectively. That is,
\begin{align}\label{3.1}
\begin{array}{l}
x\oplus y=M_{\oplus}xy,\\
x\otimes y=M_{\odot}xy,\quad x,y\in \D_k.\\
\end{array}
\end{align}

As a convention, in a ring $R =\D_k$, we always assume $\d_k^1=\vec{\bf 1}$ and $\d_k^k=\vec{\bf 0}$.

In vector form, denote by $M_{\neg}$ the structure matrix of $\neg$, then $X\oplus (\neg X)=0$ can be expressed in vector form as
\begin{align}\label{3.3}
M_{\oplus}x(M_{\neg} x)=\d_k^k,\quad x\in \D_k.
\end{align}
Similarly, $X \ominus Y$ can be expressed as
\begin{align}\label{3.4}
\begin{array}{ccl}
M_{\ominus} xy&=& M_{\oplus}xM_{\neg}y=M_{oplus}\left(I_k\otimes M_{\neg}\right)xy,\\
~&~& \quad x,y\in \D_k.
\end{array}
\end{align}

Note that in a ring, $\neg$ and $\ominus$ are not independent operators. They are uniquely determined by $\oplus$. But for computational ease, usually we also define $M_{\neg}$ and use it in the ASSR representation of networks over finite rings.

\begin{thm}\label{t.3.1} Given a finite set $R$ (where $|R|=k$) with two binary operators $\oplus$ and $\odot$, then the properties of $R$ can be verified by the corresponding properties of its operators. That is,
\begin{itemize}
\item[(i)] (Commutativity of $\oplus$): $(R,\oplus)$ is commutative, if and only if,
\begin{align}\label{3.4}
M_{\oplus} =M_{\oplus}W_{[k,k]}.
\end{align}
\item[(ii)] (Associativity of $\oplus$): $(R,\oplus)$ is associative, if and only if,
\begin{align}\label{3.5}
M_{\oplus}^2 =M_{\oplus}\left(I_k\otimes M_{\oplus}\right).
\end{align}
\item[(iii)] (Identity of $\oplus$): $\d_k^k$ is the identity for $(R,\oplus)$, if and only if,
\begin{align}\label{3.6}
M_{\oplus}\d_k^k=M_{\oplus}W_{[k,k]}\d_k^k =I_k.
\end{align}
\item[(iv)] (Invertibility of $\oplus$): $(R,\oplus)$ is invertible, if and only if, there exists a unary operator $\neg$, such that
\begin{align}\label{3.7}
M_{\oplus}\left(I_k\otimes M_{\neg}\right)PR_k=\d_k\otimes \J_k^T.
\end{align}
\item[(v)] (Associativity of $\odot$): $(R,\odot)$ is associative, if and only if,
\begin{align}\label{3.8}
M_{\odot}^2 =M_{\odot}\left(I_k\otimes M_{\odot}\right).
\end{align}
 \item[(vi)] (Identity of $\odot$): $\d_k^1$ is the identity for $(R,\odot)$, if and only if,
\begin{align}\label{3.9}
M_{\odot}\d_k^1=M_{\odot}W_{[k,k]}\d_k^1=I_k.
\end{align}
\item[(vii)] (Distributivity of $\oplus$ with respect to $\odot$): $(R,\oplus,\odot)$ is distributive, if and only if,
\begin{align}\label{3.10}
\begin{array}{lr}
M_{\odot}M_{\oplus}=M_{\oplus}M_{\odot}\left(I_{k^2}\otimes M_{\odot}\right)&~\\
~~~\left(I_{k}\otimes W_{[k,k]}\right)\left(I_{k^2}\otimes \PR_{k}\right),&(a)\\
M_{\odot}\left(I_{k}\otimes M_{\oplus}\right)=M_{\oplus}M_{\odot}\left(I_{k^2}\otimes M_{\odot}\right)&~\\
~~~\left(I_{k}\otimes W_{[k,k]}\right)\PR_{k}.&(b)\\
\end{array}
\end{align}
\end{itemize}
\end{thm}

\noindent{\it Proof}. We prove (\ref{3.10}) (b) only. The proofs for other equations are similar.
To this end, we need only to show that the (\ref{3.10}) (b) is equivalent to
\begin{align}\label{3.11}
X\odot (Y\oplus Z)=(X\odot Y)\oplus (X\odot Z),\quad X,Y,Z\in R.
\end{align}

Expressing both sides of (\ref{3.11}) into vector form, we have
$$
LHS=M_{\odot}xM_{\oplus}yz=M_{\odot}\left(I_k\otimes M_{\oplus}\right)xyz.
$$
$$
\begin{array}{ccl}
RHS&=&M_{\oplus}M_{\odot}xyM_{\odot}xz\\
~&=&M_{\oplus}M_{\odot}\left(I_{k^2}\otimes M_{\odot}\right)xyxz\\
~&=&M_{\oplus}M_{\odot}\left(I_{k^2}\otimes M_{\odot}\right)xW_{[k,k]}xyz\\
~&=&M_{\oplus}M_{\odot}\left(I_{k^2}\otimes M_{\odot}\right)\left(I_{k}\otimes W_{[k,k]}\right)x^2yz\\
~&=&M_{\oplus}M_{\odot}\left(I_{k^2}\otimes M_{\odot}\right)\left(I_{k}\otimes W_{[k,k]}\right)PR_kxyz\\
\end{array}
$$
Since $x,y,z\in \D_k$ are arbitrary, setting LHS=RHS yields (\ref{3.10}) (b). Hence, (\ref{3.10}) (b) is equivalent to (\ref{3.11}).

\hfill $\Box$

Since a finite ring $R$ is uniquely determined by its $\oplus$  and $\odot$, according to Theorem \ref{t.3.1},  when equations (\ref{3.5})-(\ref{3.10}) are satisfied, $R$ is a ring. In addition, if (\ref{3.4}) is also true, $R$ is a commutative ring.  Using it, a finite ring can be constructed by solving a set of matrix equations. We give an example for this.

\begin{exa}\label{e.3.2}~~

\begin{enumerate}
\item Consider $R=\{1,2,\cdots,p-1,0\}$, where $p$ is a primer number. Then it is easy to show that $R\simeq \Z_p$.
Let $\J$ be the identity of $R$. Then the cyclic group
$<\J>:=\{\J,2\J,3\J, \cdots,p\J=0\}$. For $n<p$ we know that $n\J\neq 0$, because if $n\J=0$ for $0<n<p$, then $g=\{\J,2\J,\cdots,n\J\}$ becomes a group, and hence $n|p$, which is a contradiction. Next, consider the product. Similar argument shows $u\J\times v\J=uv (\mbox{mod}~p)\J$. Hence, $R\cong \Z_p$.

\item Consider $R=\{1,2,3,0\}$. Assume the structures of its operators $\oplus$ and $\odot$ are $M_{\oplus}$ and $M_{\odot}$ respectively. Then by exhaustive searching it is easy to verify that there are exactly 6 rings, determined by their $M_{\oplus}$ and $M_{\odot}$ respectively as:
\begin{itemize}
\item[(i)] $R_1$:
$$
\begin{array}{l}
M_{\oplus}=\d_4[3, 4, 2, 1, 4, 3, 1, 2, 2, 1, 4, 3, 1, 2, 3, 4],\\
M_{\odot}=\d_4[1, 2, 3, 4, 2, 1, 3, 4, 3, 3, 4, 4, 4, 4, 4, 4].
\end{array}
$$
\item[(ii)] $R_2$:
$$
\begin{array}{l}
M_{\oplus}=\d_4[4, 3, 2, 1, 3, 4, 1, 2, 2, 1, 4, 3, 1, 2, 3, 4],\\
M_{\odot}=\d_4[1, 2, 3, 4, 2, 1, 3, 4, 3, 3, 4, 4, 4, 4, 4, 4].
\end{array}
$$
\item[(iii)] $R_3$:
$$
\begin{array}{l}
M_{\oplus}=\d_4[4, 3, 2, 1, 3, 4, 1, 2, 2, 1, 4, 3, 1, 2, 3, 4],\\
M_{\odot}=\d_4[1, 2, 3, 4, 2, 2, 4, 4, 3, 4, 3, 4, 4, 4, 4, 4].
\end{array}
$$
\item[(iv)] $R_4$:
$$
\begin{array}{l}
M_{\oplus}=\d_4[4, 3, 2, 1, 3, 4, 1, 2, 2, 1, 4, 3, 1, 2, 3, 4],\\
M_{\odot}=\d_4[1, 2, 3, 4, 2, 3, 1, 4, 3, 1, 2, 4, 4, 4, 4, 4].
\end{array}
$$
\item[(v)] $R_5$:
$$
\begin{array}{l}
M_{\oplus}=\d_4[2, 3, 4, 1, 3, 4, 1, 2, 4, 1, 2, 3, 1, 2, 3, 4],\\
M_{\odot}=\d_4[1, 2, 3, 4, 2, 4, 2, 4, 3, 2, 1, 4, 4, 4, 4, 4].
\end{array}
$$
\item[(vi)] $R_6$:
$$
\begin{array}{l}
M_{\oplus}=\d_4[4, 3, 2, 1, 3, 4, 1, 2, 2, 1, 4, 3, 1, 2, 3, 4],\\
M_{\odot}=\d_4[1, 2, 3, 4, 2, 4, 2, 4, 3, 2, 1, 4, 4, 4, 4, 4].
\end{array}
$$
%
\end{itemize}
\end{enumerate}
\end{exa}

\begin{rem}\label{r.3.3}
From Example \ref{e.3.2}, the  following facts are observable easily:
\begin{itemize}
\item[(1)] All $R_i$, $i=1,2,3,4,5,6$ obtained in Example \ref{e.3.2} are commutative.
\item[(2)] $R_5=\Z_4$.
\item[(3)] Among them $R_1$ and $R_5$ are automorphic. The automorphism is $\pi: (1,2,3,0)\mapsto (1,3,2,0)$.
\end{itemize}

It can also be seen from Example \ref{e.3.2} that if $s$ is not a primer number, $R=\Z_s$ is not the only commutative ring for $|R|=s$.
\end{rem}

\section{Properties of Networks Over Finite Rings}

In general, a network over a finite ring $R$  can be expressed as
\begin{align}\label{4.1}
\begin{cases}
X_1(t+1)=p_1(X_1,X_2,\cdots,X_n),\\
X_2(t+1)=p_2(X_1,X_2,\cdots,X_n),\\
\vdots,\\
X_n(t+1)=p_n(X_1,X_2,\cdots,X_n),\\
\end{cases}
\end{align}
where $X_i\in R$, $i\in [1,n]$, and
$$
\begin{array}{ccl}
p_i&:=&\dsum_{j\in \Lambda_i}a^i_{j}X_1^{r^1_j}X_2^{r^2_j}\cdots X_n^{r^n_j},\\
~&~&\quad a^i_j,X_j\in R,\; j\in \Lambda_i, i\in [1,n],
\end{array}
$$
are polynomials, $\Lambda_i=[1,\ell_i]$, $i\in [1,n]$, are index sets with $\ell_i$ as the number of terms in $p_i$.

Similarly, a control network over a finite ring $R$  can be expressed as
\begin{align}\label{4.2}
\begin{array}{l}
\begin{cases}
X_1(t+1)=p_1(X_1,\cdots,X_n,U_1,\cdots,U_m),\\
X_2(t+1)=p_2(X_1,\cdots,X_n,U_1,\cdots,U_m),\\
\vdots,\\
X_n(t+1)=p_n(X_1,\cdots,X_n,U_1,\cdots,U_m)),\\
\end{cases}\\
Y_j(t)=\xi_j(X_1,\cdots,X_n),\quad j\in [1,p].
\end{array}
\end{align}
where $X_i,U_s,Y_j\in R$,  $p_i$, $\xi_j$ are polynomials, $i\in [1,n]$, $s\in[1,m]$, and $j\in[1,p]$.

Note that for notational ease, in a (control) network dynamic equation the product symbol $\odot$ is omitted, the addition $\oplus$, negation $\neg$ (or subtraction $\ominus$) are simply replaced by $+$ and $-$ respectively. But you have to keep this fact in mind that all the operators are determined by the two operators $\oplus$ and $\odot$ of the bearing finite ring.

We give an example to depict this.

\begin{exa}\label{e4.1} Consider a control network $\Sigma$ over $R$. Its network graph is depicted in Figure \ref{Fig.4.1}, and its dynamic equation is described by the following (\ref{4.3}).

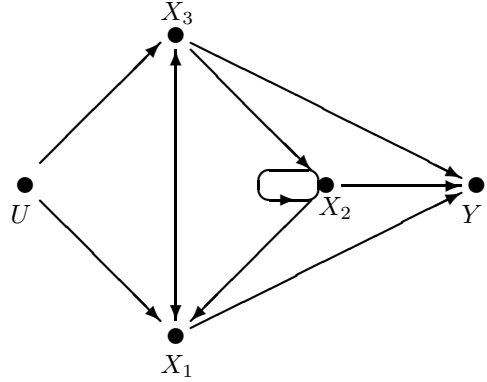
\begin{figure}
\centering
\setlength{\unitlength}{1cm}
\begin{picture}(8,6)\thicklines
\put(1,3){\circle*{0.2}}
\put(3,1){\circle*{0.2}}
\put(3,5){\circle*{0.2}}
\put(5,3){\circle*{0.2}}
\put(7,3){\circle*{0.2}}
\put(1.2,2.8){\vector(1,-1){1.6}}
\put(1.2,3.3){\vector(1,1){1.6}}
\put(3,1.2){\vector(0,1){3.6}}
\put(3,4.8){\vector(0,-1){3.6}}
\put(4.8,2.8){\vector(-1,-1){1.6}}
\put(3.2,4.8){\vector(1,-1){1.6}}
\put(3.2,4.9){\vector(2,-1){3.6}}
\put(3.2,1.1){\vector(2,1){3.6}}
\put(5.2,3){\vector(1,0){1.6}}
\put(4.5,3){\oval(0.8,0.4)}
\put(4.4,2.8){\vector(1,0){0.2}}
\put(0.8,2.5){$U$}
\put(2.8,0.5){$X_1$}
\put(2.8,5.2){$X_3$}
\put(4.9,2.6){$X_2$}
\put(6.8,2.5){$Y$}

\end{picture}
\caption{Network Graph of (\ref{4.3})\label{Fig.4.1}}
\end{figure}

\begin{align}\label{4.3}
\begin{array}{l}
\begin{cases}
X_1(t+1)=(X_2(t)+X_3(t)^2)U(t)\\
X_2(t+1)=X_3(t)-X_2(t)\\
X_3(t+1)=(X_1(t)+U(t))^2,\\
\end{cases}\\
Y(t)=X_1(t)+X_2(t)+X_3(t).
\end{array}
\end{align}

This example will be considered again later.
\end{exa}

It is obvious that a network over a finite ring $R$ with $|R|=k$ is a special case of $k$-valued logic. Hence the technique developed for $k$-valued logic is applicable to (control) networks over finite rings.

Before reconsider the control network (\ref{4.3}), we first consider a network over a finite ring.

\begin{exa}\label{e.4.1} Consider  $R=\Z_5$, and a network on $R$ is

\begin{align}\label{4.4}
\begin{cases}
X_1(t+1)=(X_2(t)+X_3(t))^2\\
X_2(t+1)=-X_3(t)\\
X_3(t+1)=X_1(t)-X_2(t)^2.
\end{cases}
\end{align}
Let $M_p=M_{+_5}$, $M_t=M_{\times_5}$, $M_s=M_{-_5}$, $M_m=M{\neg_5}$. Then a simple calculation shows that
\begin{tiny}
\begin{align}\label{4.5}
\begin{array}{ccl}
M_p&=&\d_5[2,3,4,5,1,3,4,5,1,2,4,5,1,2,3,5,1,2,3,4,1,2,3,4,5];\\
M_t&=&\d_5[1,2,3,4,5,2,4,1,3,5,3,1,4,2,5,4,3,2,1,5,5,5,5,5,5];\\
M_s&=&\d_5[5,4,3,2,1,1,5,4,3,2,2,1,5,4,3,3,2,1,5,4,4,3,2,1,5];\\
M_n&=&\d_5[4,3,2,1,5].
\end{array}
\end{align}
\end{tiny}
Using the structure matrices $M_p$, $M_t$, and $M_m$, we can calculate that
\begin{align}\label{4.6}
\begin{array}{ccl}
x_1(t+1)&=&M_tM_px_2(t)x_3(t)M_px_2(t)x_3(t)\\
~&=&M_tM_p(I_{25}\otimes M_p)(x_2(t)x_3(t))^2\\
~&=&M_tM_p(I_{25}\otimes M_p)PR_{25}x_2(t)x_3(t)\\
~&=&M_tM_p(I_{25}\otimes M_p)PR_{25}(\J_5^T\otimes I_{25})x(t)\\
~&:=&M_1x(t),
\end{array}
\end{align}
where $x(t)=\ltimes_{i=1}^3x_i(t)$,
\begin{tiny}
$$
\begin{array}{ccl}
M_1&=\d_{5}[&4,4,1,5,1,4,1,5,1,4,1,5,1,4,4,5,1,4,4,1,1,4,4,1,5,\\
  ~&       ~&4,4,1,5,1,4,1,5,1,4,1,5,1,4,4,5,1,4,4,1,1,4,4,1,5,\\
  ~&       ~&4,4,1,5,1,4,1,5,1,4,1,5,1,4,4,5,1,4,4,1,1,4,4,1,5,\\
  ~&       ~&4,4,1,5,1,4,1,5,1,4,1,5,1,4,4,5,1,4,4,1,1,4,4,1,5,\\
  ~&       ~&4,4,1,5,1,4,1,5,1,4,1,5,1,4,4,5,1,4,4,1,1,4,4,1,5].\\
\end{array}
$$
\end{tiny}
\begin{align}\label{4.7}
\begin{array}{ccl}
x_2(t+1)&=&M_mx_3(t)\\
~&=&(\J_{25}\otimes I_5)x(t)\\
~&:=&M_2x(t),\\
\end{array}
\end{align}
where
\begin{tiny}
$$
\begin{array}{ccl}
M_2&=\d_5[&4,3,2,1,5,4,3,2,1,5,4,3,2,1,5,4,3,2,1,5,4,3,2,1,5,\\
  ~&     ~&4,3,2,1,5,4,3,2,1,5,4,3,2,1,5,4,3,2,1,5,4,3,2,1,5,\\
  ~&     ~&4,3,2,1,5,4,3,2,1,5,4,3,2,1,5,4,3,2,1,5,4,3,2,1,5,\\
  ~&     ~&4,3,2,1,5,4,3,2,1,5,4,3,2,1,5,4,3,2,1,5,4,3,2,1,5,\\
  ~&     ~&4,3,2,1,5,4,3,2,1,5,4,3,2,1,5,4,3,2,1,5,4,3,2,1,5].\\
\end{array}
$$
\end{tiny}
\begin{align}\label{4.8}
\begin{array}{ccl}
x_3(t+1)&=&M_px_1(t)M_mM_tx_2(t)x_2(t)\\
~&=&M_px_1(t)M_mM_tPR_5x_2(t)\\
~&=&M_p(I_{5}\otimes (M_mM_tPR_5))x_1(t)x_2(t)\\
~&=&M_p(I_{5}\otimes (M_mM_tPR_5))(I_25\otimes \J_5^T)x(t)\\
~&:=&M_3x(t),\\
\end{array}
\end{align}
where
\begin{tiny}
$$
\begin{array}{ccl}
M_3&=\d_5[&5,5,5,5,5,2,2,2,2,2,2,2,2,2,2,5,5,5,5,5,1,1,1,1,1,\\
  ~&     ~&1,1,1,1,1,3,3,3,3,3,3,3,3,3,3,1,1,1,1,1,2,2,2,2,2,\\
  ~&     ~&2,2,2,2,2,4,4,4,4,4,4,4,4,4,4,2,2,2,2,2,3,3,3,3,3,\\
  ~&     ~&3,3,3,3,3,5,5,5,5,5,5,5,5,5,5,3,3,3,3,3,4,4,4,4,4,\\
  ~&     ~&4,4,4,4,4,1,1,1,1,1,1,1,1,1,1,4,4,4,4,4,5,5,5,5,5].\\
\end{array}
$$
\end{tiny}
Finally, the ASSR of (\ref{4.4}) is
\begin{align}\label{4.9}
x(t+1)=Mx(t),
\end{align}
where
\begin{tiny}
$$
\begin{array}{ccll}
M&=&~&M_1*M_2*M_3\\
~&=&\d_{125}[& 95, 90, 10,105, 25, 92, 12,107,  2, 97,\\
~&~&         ~& 17,112,  7, 77, 97,120, 15, 85, 80, 25,\\
~&~&         ~& 16, 86, 81,  1,121, 91, 86,  6,101, 21,\\
~&~&         ~& 93, 13,108,  3, 98, 18,113,  8, 78, 98,\\
~&~&         ~&116, 11, 81, 76, 21, 17, 87, 82,  2,122,\\
~&~&         ~& 92, 87,  7,102, 22, 94, 14,109,  4, 99,\\
~&~&         ~& 19,114,  9, 79, 99,117, 12, 82, 77, 22,\\
~&~&         ~& 18, 88, 83,  3,123, 93, 88,  8,103, 23,\\
~&~&         ~& 95, 15,110,  5,100, 20,115, 10, 80,100,\\
~&~&         ~&118, 13, 83, 78, 23, 19, 89, 84,  4,124,\\
~&~&         ~& 94, 89,  9,104, 24, 91, 11,106,  1, 96,\\
~&~&         ~& 16,111,  6, 76, 96,119, 14, 84, 79, 24,\\
~&~&         ~& 20, 90, 85,  5,125].\\
\end{array}
$$
\end{tiny}
Using ASSR of a network, all the properties of the network can be revealed. For instance, to find  fixed points of (\ref{4.4}), it is ready to calculate that
$$
\tr(M)=2.
$$
 Hence, there are two fixed points. Using standard technique for $k$-valued logical networks \cite{che11}, it is easy to  calculate that
$$
x^1=\d_{125}^{104}=\d_5^5\ltimes \d_5^1\ltimes \d_5^4\sim (0,1,4).
$$
That is $x_1=0$, $x_2=1$, and $x_3=4$ is a fixed point.

$$
x^2=\d_{125}^{125}=\d_5^5\ltimes \d_5^5\ltimes \d_5^5\sim (0,0,0).
$$
That is $x_1=0$, $x_2=0$, and $x_3=0$ is another fixed point.
\end{exa}

Next, we consider the control problems of control networks over finite rings.
We use control network (\ref{4.3}) to describe this.

\begin{exa}\label{e.4.2} Recall control network (\ref{4.3}).  Its bearing ring is assumed to be
$R_2$ in Example \ref{e.3.2}. Then the structure matrices of $\oplus$, $\odot$, and $\neg$, denoted by $M_P$, $M_t$, and $M_n$ respectively, are as follows:
$$
\begin{array}{l}
M_p=\d_4[3, 4, 2, 1, 4, 3, 1, 2, 2, 1, 4, 3, 1, 2, 3, 4],\\
M_t=\d_4[1, 2, 3, 4, 2, 1, 3, 4, 3, 3, 4, 4, 4, 4, 4, 4],\\
M_n=\d_4[2,1,3,4].
\end{array}
$$
To get the ASSR of (\ref{4.3}) we have
$$
\begin{array}{ccl}
x_1(t+1)&=&M_tM_px_2(t)M_tx^2_3(t)u(t)\\
~&=&M_tM_px_2(t)M_t\PR_4x_3(t)u(t)\\
~&=&M_tM_p(I_4\otimes (M_t\PR_4))x_2(t)x_3(t)u(t)\\
~&=&M_tM_p(I_4\otimes (M_t\PR_4))(\J^T_4\otimes I_{64})x(t)u(t)\\
~&=&M_tM_p(I_4\otimes (M_t\PR_4))\\
~&~&(\J^T_4\otimes I_{64})W_{[4,64]}u(t)x(t)\\
~&:=&L_1u(t)x(t),
\end{array}
$$
where
$$
L_1=\d_4[3,3,1,1,\cdots,4,4,4,4]\in {\cal L}_{4\times 256}.
$$

$$
\begin{array}{ccl}
x_2(t+1)&=&M_px_3(t)M_mx_2(t)\\
~&=&M_p(I_4\otimes M_m)x_3(t)x_2(t)\\
~&=&M_p(I_4\otimes M_m)W_{[4,4]}x_2(t)x_3(t)\\
~&=&M_p(I_4\otimes M_m)W_{[4,4]}(\J^T_{16}\otimes I_{16})u(t)x(t)\\
~&:=&L_2u(t)x(t),
\end{array}
$$
where
$$
L_2=\d_4[4,3,1,2,\cdots,1,2,3,4]\in {\cal L}_{4\times 256}.
$$

$$
\begin{array}{ccl}
x_3(t+1)&=&M_tM_pu(t)x_1(t)M_pu(t)x_1(t)\\
~&=&M_tM_p(I_{16}\otimes M_p)(u(t)x_1(t))^2\\
~&=&M_tM_p(I_{16}\otimes M_p)\PR_{16}u(t)x_1(t)\\
~&=&M_tM_p(I_{16}\otimes M_p)\PR_{16}\\
~&~&(I_{16}\otimes \J^T_{16})u(t)x(t)\\
~&:=&L_3u(t)x(t),
\end{array}
$$
where
$$
\begin{array}{ccl}
L_3&=&\d_4[4,4,4,4,\cdots,4,4,4,4]\\
~&\in& {\cal L}_{4\times 256}.
\end{array}
$$
$$
y(t)=M_pM_px_1(t)x_2(t)x_3(t):=Ex(t),
$$
where
$$
E=\d_4[1,1,4,3,\cdots,1,2,3,4]\in {\cal L}_{4\times 64}.
$$.

Finally, we have the ASSR of (\ref{4.3}) as
\begin{align}\label{4.11}
\begin{cases}
x(t+1)=Lu(t)x(t),\\
y=Ex,
\end{cases}
\end{align}
where
$$
\begin{array}{ccl}
L&=&L_1*L_2*L_3=\d_{256}[48,84,4,8,\cdots,52,56,60,64]\\
~&\in& {\cal L}_{256\times 256}.
\end{array}
$$

Using ASSR (\ref{4.11}), all classical control problems can be solved. For instance, we consider
\begin{itemize}
\item[(i)] Controllability:
Set
$$
\begin{array}{ccl}
M&:=&{\dsum_{{\cal B}}}_{i=1}^4L\d_4^i\\
~&=&\begin{bmatrix}
0&0&0&0&\cdots&1&0&0&0\\
0&0&0&0&\cdots&0&0&0&0\\
\vdots&~&~&~&~&~&~&~&~\\
0&0&0&0&\cdots&0&0&0&1\\
\end{bmatrix}\in {\cal B}_{64\times 64}.
\end{array}
$$
Then we know that
$$
\tr(M)=2.
$$
Moreover,
$$
M(60,60)=M(64,64)=1.
$$
Hence, there are two control fixed points: $\d_{64}^{60}\sim(4,3,4)$ and $\d_{64}^{64}\sim(4,4,4)$\footnote{
A point $x_0$ is said to be a control fixed point, if there exists a control $u_0$ such that
$x_0=Lu_0x_0$. \cite{zha20}.}

Using $M$, the controllability matrix can be calculated as
$$
\begin{array}{ccl}
{\cal C}&:=&{\dsum_{{\cal B}}}_{i=1}^{64}M^{(i)}\\
~&=&\begin{bmatrix}
1&1&1&1&\cdots&1&1&1&1\\
0&0&0&0&\cdots&0&0&0&0\\
\vdots&~&~&~&~&~&~&~\\
1&1&1&1&\cdots&1&1&1&1\\
\end{bmatrix}\in {\cal B}_{64\times 64}.
\end{array}
$$
Hence the system is not completely controllable.

It is easy to see that
$$
\Row_j({\cal C})=\J_{64}, \quad j\in J,
$$
where
$$
J=\{1,5,17,21,36,40,41,48,52,56,57,60,61,64\}.
$$
Hence $\d_{64}^j$, $j\in J$ are globally reachable. That is, starting from any initial point $x_0$, there exists a sequence of controls $u(t),t\geq 0$, which can drive the trajectory from $x_0$ to $\d_{64}^j$, $j\in J$.

\item[(ii)] Stabilization: Stabilizing to $x_d$  means for any initial point $x_0$ there exits a sequence of controls $u(t),~t\geq 0$ and a $T>0$, such that the controlled trajectory $x(t)$ with $x(0)=x_0$ satisfies $x(t)=x_d$, $t>T$. It is easy to see that a network is stabilizable to $x_d$, if and only if, the system is globally reachable to $x_d$ and $x_d$ is a control fixed point.

According to (i), the network (\ref{4.7}) is stabilizable to $\d_{64}^{60}\sim (4,3,4)$ and $\d_{64}^{64}\sim (4,4,4)$.

\item[(iii)] Synchronization:

Synchronization means the network can reach a point $x_d=(x^d_1,x^d_2,\cdots,x^d_n)$ with $x^d_i=x^d_0$, $i\in[1,n]$ for $t>T$ \cite{men20}. The control synchronization means under certain control the controlled network can reach synchronization.

Consider control network (\ref{4.11}). It follows from the aforementioned argument that it can be synchronized by a sequence of controls to the position $(4,4,4)$.

\end{itemize}
\end{exa}

A last comment is: all the techniques developed for logical (control) networks are applicable to networks over finite rings.

Next, let's see what is new for networks over finite rings.

\section{Sub-networks over Ideals}

Hereafter, we consider only finite commutative rings. The ``commutativity" is always assumed and will not be mentioned repeatedly.

Consider a finite ring $R$. Let $S\subset R$ be an ideal. Since $S$ itself is a ring, there exist a ring $R_S$, called the essential ring of $S$, and an isomorphism $\pi: S\ra R_S$.

Now for each $a\in R$, since $S$ is an idea, then for each $s_i\in S$ there exists a unique $s_{j}\in S$ such that
\begin{align}\label{5.1.1}
as_i=s_{j},\quad s_i,s_j\in S.
\end{align}
That is, each $a\in R$ can induce a mapping $S\ra S$, denoted by $\theta_a$. Now let $R_S:=\{1,2,\cdots,k-1,0\}$, where $k=|S|$. Then in vector form $\theta_a$ has its structure matrix as $\Theta_a$, such that
\begin{align}\label{5.1.2}
\pi(as_i)=\Theta_a\pi(s_i).
\end{align}

We use an example to explain this.

\begin{exa}\label{e.5.1.1} Consider $\Z_6$, and $S=\{3,0\}\subset \Z_6$. It is easy to see that
\begin{itemize}
\item[(i)]
$S$ is an ideal of $\Z_6$.
\item[(ii)] $R_S=\Z_2$, and the isomorphism is
$$
\pi(3)=1,\quad \pi(0)=0.
$$
\item[(iii)] Since $1\times_6 3=3$, $1\times_6 6=6$,
$
\Theta_1=\d_2[1,2].
$

Similarly, we have
\begin{align}\label{5.1.3}
\begin{array}{l}
\Theta_1=\Theta_3=\Theta_5=I_2,\\
\Theta_2=\Theta_4=\Theta_6=\d^2[2,2].\\
\end{array}
\end{align}
\end{itemize}
\end{exa}

\begin{dfn}\label{d5.1.100} Let $R$ be a finite ring and $S\subset R$ be an ideal. $S$ is called a proper idea if there exists a mapping $\phi: R\ra S$ such that
\begin{align}\label{5.1.301}
\Theta_a(s)=\phi(a)s,\quad a\in R, s\in S.
\end{align}
\end{dfn}

\begin{exa}\label{e.5.1.101}
\begin{itemize}
\item[(i)] Consider $\Z_6$ and $S=\{3,0\}\subset \Z_6$. $S$ is a proper ideal of $\Z_6$, because
according to (\ref{5.1.3})
we have
\begin{align}\label{5.1.3a}
\begin{array}{l}
\Theta_1(x)=\Theta_3(x)=\Theta_5(x)=3x\\
\Theta_2(x)=\Theta_4(x)=\Theta_0(x)=0,\\
\end{array}
\end{align}
which leads to
\begin{align}\label{5.1.3b}
\begin{array}{l}
\phi(1)=\phi(3)=\phi(5)=3\\
\phi(2)=\phi(4)=\phi(0)=0.\\
\end{array}
\end{align}

\item[(ii)]
Let $R=\left(\{1,2,\cdots,\kappa-1,0\},\oplus,\odot\right)$ be a finite ring and $S\subset R$ be an ideal of $R$ with $R_S=\Z_r$.

Assume the elements in $R$ satisfies the following (\ref{5.1.300}).
\begin{align}\label{5.1.300}
a=\underbrace{1\oplus 1\oplus \cdots \oplus 1}_a,
\end{align}

Then we show that $S$ is a proper ideal of $R$.

Define $\phi:R\ra S$ as
\begin{align}\label{5.1.3c}
\phi(a)= a~(\mbox{mod}~r).
\end{align}
Let $\pi:S\ra \Z_r$ be the isomorphism and set
$$
\xi=\pi^{-1}({\bf 1}).
$$
Then it is clear that
$$
S=\{\xi,2\xi,\cdots,(r-1)\xi,0\}.
$$
Now for any $a\in R$,
\begin{align}\label{5.1.3011}
\begin{array}{ccl}
\pi(as)&=&\pi(at\xi)=\pi[(\a r+\b)t\xi]\\
~&=&=\pi(\b\xi) t=\b t,
\end{array}
\end{align}
where $\b=a~(\mbox{mod}~r)$. The last equality of
(\ref{5.1.3011}) comes from the fact that  $rt\xi=0$.
Hence the $\phi$ defined in (\ref{5.1.3c}) satisfies (\ref{5.1.301}).

Note that assumption (\ref{5.1.300}) is not always true. Recall the six rings obtained in Example \ref{e.3.2}, only $R_5$, which is $\Z_4$, satisfies it. Hence, (\ref{5.1.301}) is not always true.

\end{itemize}
\end{exa}

\begin{dfn}\label{p.5.1.102} Let $R=\{1,2,\cdots,\kappa-1,0\}$ be a finite ring and $S\subset R$ be a proper ideal with $|S|=r$.

Consider  a network $\Sigma$  over $R$, denoted by
\begin{align}\label{5.1.4}
\begin{array}{ccl}
Z_i(t+1)&=&p_i(Z_1(t),Z_2(t),\cdots,Z_n(t))\\
~&=& \dsum_{j=1}^{\ell_i}a^i_jZ_1^{r^1_j}\cdots Z_n^{r^n_j},\quad i\in[1,n].
\end{array}
\end{align}

We construct a network on $R_s$, called the sub-network over $S$, as
\begin{align}\label{5.1.5}
X_i(t+1)=\dsum_{j=1}^{\ell_i}\phi(a^i_j)X_1^{r^1_j}\cdots X_n^{r^n_j},\quad
i\in[1,k].
\end{align}
\end{dfn}

Then we have the following result:

\begin{prp}\label{p.5.1.2} Consider the network $\Sigma$ over $R$ defined by (\ref{5.1.4}). Let $S\subset R$ be a proper ideal. Then for each $s_0\in S$, $i=1,2,\cdots,n$. The trajectory of $\Sigma$ on $R$ stating from $s_0$, denoted by $z(t,s_0)$, $t\geq 0$, and the corresponding trajectory of $\Sigma_S$ on $R_S$, denoted by $x(t,\phi(s_0))$, $t\geq 0$, the following identity holds:
\begin{align}\label{5.1.6}
\phi\left(z(t,s_0)\right)=x(t,\phi(s_0)),\quad t\geq 0.
\end{align}
\end{prp}

\noindent{\it Proof}.
By construction of $\Sigma_S$ it is clear that
$$
\phi(\Sigma|_S)=\Sigma_S.
$$
Since $\phi$ is an isomorphism, for any polynomial $p$ we have
$$
\phi[p(X_1,X_2,\cdots,X_n)]=\phi(p)(\pi(X_1),\phi(X_2),\cdots,\phi(X_n)),
$$
where $X_i\in S$, $i=1,2,\cdots,n$, and $\phi(p)$ is defined exactly as in (\ref{5.1.5}).

Assume $Z(t)=(Z_1(t),\cdots,Z_n(t))\in S$, then
$$
Z(t+1)=\Sigma|_SZ(t).
$$
Hence
$$
\begin{array}{ccl}
\phi(Z(t+1))&=&\phi[\Sigma|_SZ(t)]\\
&=&\pi(\Sigma|_S)\phi(Z(t))\\
=\Sigma_SX(t).
\end{array}
$$
The conclusion follows.

\hfill $\Box$

\begin{exa}\label{e.5.1.3} Consider $R=\Z_6$. An network $\Sigma$ is defined by
\begin{align}\label{5.1.7}
\begin{cases}
Z_1(t+1)=4Z^2_1(t)-Z_2(t),\\
Z_2(t+1)=Z_1(t)Z_2(t),\\
\end{cases}
\end{align}

\begin{align}\label{5.1.701}
\begin{array}{ccl}
M_{+_6}&=&\d_6[2,3,4,5,6,1,3,4,5,6,1,2,\\
~&~      &~~~~~4,5,6,1,2,3,5,6,1,2,3,4,\\
~&~      &~~~~~6,1,2,3,4,5,1,2,3,4,5,6],\\
M_{\times_6}&=&\d_6[1,2,3,4,5,6,2,4,6,2,4,6,\\
~&~           &~~~~~3,6,3,6,3,6,4,2,6,4,2,6,\\
~&~           &~~~~~5,4,3,2,1,6,6,6,6,6,6,6],\\
M_{\neg_6}&=&\d_6[5,4,3,2,1,6].
\end{array}
\end{align}

Its ASSR can be calculated as follows:
$$
\begin{array}{l}
z_1(t+1)=M_{+_6}(M_{\times_6}\d_6^4)M_{\times_6}\PR_6z_1(t)M_{\neg_6}z_2(t)\\
~~~=M_{+_6}(M_{\times_6}\d_6^4)M_{\times_6}\PR_6\left(I_6\otimes M_{\neg_6}\right)z(t)\\
~~~:=M_1z(t),
\end{array}
$$
where
$$
\begin{array}{ccl}
M_1&=\d_6[&3,2,1,6,5,4,3,2,1,6,5,4,\\
~&~&       5,4,3,2,1,6,3,2,1,6,5,4,\\
~&~&       3,2,1,6,5,4,5,4,3,2,1,6].
\end{array}
$$

$$
z_2(t+1)=M_{\times_6}z(t):=M_2z(t),
$$
where
$$
\begin{array}{ccl}
M_2&=&M_{\times_6}\\
~&=\d_6[&1,     2,     3,     4,     5,     6,     2,     4,     6,     2,     4,     6, \\
~&~&      3,     6,     3,     6,     3,     6,     4,     2,     6,     4,2,     6,  \\
~&~&   5,     4,     3,     2,     1,     6,     6,     6,     6,     6,     6,     6].
\end{array}
$$

Finally, the ASSR of (\ref{5.1.7}) is
\begin{align}\label{5.1.8}
z(t+1)=Mz(t),
\end{align}
where
$$
\begin{array}{ccl}
M&=&M_1*M_2\\
~&=\d_{36}&13,     8,     3,    34,    29,    24,    14,    10,     6,    32,    28,    24,\\
~&~&       27,    24,    15,    12,     3,    36,    16,     8,     6,    34,    26,    24,\\
~&~&       17,    10,     3,    32,    25,    24,    30,    24,    18,    12,     6,    36].
\end{array}
$$

Next, consider $S=\{3,0\}$ as an ideal of $\Z_6$. We construct $\Sigma_S$ as follows: Since $\theta_4=\d_2[2,2]$, i.e. $\theta_4(x)={\bf 0}_2$, the sub-network over $S$ becomes
\begin{align}\label{5.1.9}
\begin{cases}
X_1(t+1)=-X_2(t),\\
X_2(t+1)=X_1(t)X_2(t).\\
\end{cases}
\end{align}

Its ASSR is calculated as follows:
$$
\begin{array}{ccl}
x_1(t+1)&=&M_{\neg_2}x_2(t)=M_{\neg_2}\left(\J_2\otimes I_2\right)x(t)\\
~&:=&N_1x(t),
\end{array}
$$
where
$$
N_1=\d_2[  1,     2,     1,     2];
$$
$$
x_2(t+1)=M_{\times_2}x(t):=N_2x(t),
$$
where
$$
N_2=M_{\times_2}
=\d_2[1,2,2,2].
$$

Finally we have
\begin{align}\label{5.1.10}
x(t+1)=Nx(t),
\end{align}
where
$$
N=N_1*N_2=\d_4[1,4,2,4].
$$

Now we are ready to verify the consistency of corresponding trajectories:

\begin{itemize}
\item[(i)] Assume $Z_0=(3,3)$, then it is easy to verify that $z_0=\d_{36}^{15}$ is a fixed point of (\ref{5.1.8}).
Let $X_0=\pi(Z_0)=(1,1)$. Then it is ready to verify that $x_0=\d_{4}^{1}$ is a fixed point of (\ref{5.1.10}).
\item[(ii)] Assume $Z_0=(3,6)$, then  $z_0=\d_{36}^{18}$. The trajectory of (\ref{5.1.8})
is
$$z(t,z_0)=\{\d_{36}^{18}, \d_{36}^{36}, \d_{36}^{36},\cdots\}.
$$
Let $X_0=\pi(Z_0)=(1,0)$. Then $x_0=\d_{4}^{2}$. The trajectory  of (\ref{5.1.10}) is
$$x(t,x_0)=\{\d_{4}^{2}, \d_{4}^{4}, \d_{4}^{4},\cdots\}.
$$
It is ready to verify that $\pi(z(t,z_0,t))=x(t,x_0)$.
\item[(iii)] Assume $Z_0=(6,3)$, then  $z_0=\d_{36}^{33}$. The trajectory of (\ref{5.1.8})
is
$$z(t,z_0)=\{\d_{36}^{33},\d_{36}^{18}, \d_{36}^{36}, \d_{36}^{36},\cdots\}.
$$
Let $X_0=\pi(Z_0)=(0,1)$. Then $x_0=\d_{4}^{3}$. The trajectory  of (\ref{5.1.10}) is
$$x(t,x_0)=\{\d_4^3,\d_{4}^{2}, \d_{4}^{4}, \d_{4}^{4},\cdots\}.
$$
We also have $\pi(z(t,z_0))=x(t,x_0)$.
\item[(iv)] Assume $Z_0=(6,6)$ and $X_0=\pi(Z_0)=(0,0)$. Then it is easy to verify that $z_0=\d_{36}^{36}$ and $x_0=\d_4^4$ are the fixed points of  (\ref{5.1.8}) and (\ref{5.1.10}) respectively.
\end{itemize}
\end{exa}

Next, we consider a control network  $\Sigma^C$  over $R$, denoted by
\begin{align}\label{5.1.11}
\begin{array}{l}
\begin{array}{ccl}
Z_i(t+1)&=&p_i(Z_1(t),\cdots,Z_n(t),U_1(t),\cdots,U_{m}(t))\\
~&=& \dsum_{j=1}^{\ell_i}a^i_jZ_1^{r^1_j}\cdots Z_n^{r^n_j}U_1^{s^1_j}\cdots U_m^{s^m_j},\quad i\in [1,n],
\end{array}\\
\begin{array}{ccl}
Y_{\ell}&=&\xi_{\ell}(Z_1(t),\cdots,Z_n(t))\\
~&=&\dsum_{\a=1}^{\ell_i}b^i_jZ_1^{e^1_j}\cdots Z_n^{e^n_j},\quad \ell\in [1,p].
\end{array}
\end{array}
\end{align}

Let $S\subset R$ be an ideal. Then we construct a network on $R_s$, called the sub-network over $S$, as
\begin{align}\label{5.1.12}
\begin{array}{l}
\begin{array}{ccl}
X_i(t+1)&=&\dsum_{j=1}^{\ell_i}\phi(a^i_j)X_1^{r^1_j}\cdots X_n^{r^n_j}\phi(U_1)^{s^1_j}\cdots \phi(U_m)^{s^m_j},\\
~&~&i\in [1,n],
\end{array}\\
Y_{\ell}=\dsum_{\a=1}^{\ell_i}\phi(b^i_j)X_1^{e^1_j}\cdots X_n^{e^n_j},\quad
\ell\in[1,p].
\end{array}
\end{align}

Similar to network/sub-network case, the control network/sub-network satisfies the following result.

\begin{prp}\label{p.5.1.4} Consider the control network $\Sigma^C$ over $R$ defined by (\ref{5.1.11}). Let $S\subset R$ be a proper ideal.  Then for each $s_0, u(t)\in S$, the trajectory of $\Sigma$ on $R$ stating from $s_0$ and derived by $U(t)=(U_1(t),U_2(t),\cdots,U_m(t))$, $t=0,1,2,\cdots$, denoted by $z(t,u(t),s_0)$, $t\geq 0$, and the corresponding trajectory of $\Sigma_S$ on $R_S$, stating from $\pi(s_0)$ and derived by $V(t)=\pi(U(t))$, denoted by $x(t,\pi(u(t)),\pi(s_0))$, $t\geq 0$, the following identity holds:
\begin{align}\label{5.1.13}
\pi\left(z(t,u(t),s_0)\right)=x(t,\pi(u(t)),\pi(s_0)),\quad t\geq 0.
\end{align}
\end{prp}

We also give an example to verify it.

\begin{exa}\label{e.5.1.5} Consider $R=\Z_6$. An network $\Sigma^C$ is defined by
\begin{align}\label{5.1.14}
\begin{cases}
Z_1(t+1)=4Z^2_1(t)-Z_2(t)+U(t),\\
Z_2(t+1)=Z_1(t)Z_2(t),\\
\end{cases}
\end{align}
which is obtained from (\ref{5.1.7}) be adding a control $U(t)$ to its first equation. Its reduced sub-network on $S$ is
\begin{align}\label{5.1.15}
\begin{cases}
X_1(t+1)=U(t)-X_2(t),\\
X_2(t+1)=X_1(t)X_2(t).\\
\end{cases}
\end{align}

We skip the detailed computation and give their ASSRs  as follows:
\begin{align}\label{5.1.16}
z(t+1)=Lu(t)x(t),
\end{align}
where
$$
L=\d_{36}[19,14,9,4,0,18,12,6,36]\in {\cal L}_{36\times 216}.
$$
\begin{align}\label{5.1.17}
x(t+1)=Fv(t)x(t),
\end{align}
where
$$
F=\d_{4}[3,2,4,2,1,4,2,4].
$$

Assume we choose $s_0=\d_{36}^{15}\sim (3,3)$ and choose
$$
u(t)=
\begin{cases}
\d_{6}^{3},\quad t=0,2,4,\cdots\\
\d_{6}^{6},\quad t=1,3,5,\cdots.
\end{cases}
$$

Then the trajectory on $\Z_6$ is
$$
\begin{array}{ccl}
Z_1(t,s_0)&=&\{3,6,3,3,6,3,6,3,6,\cdots\},\\
z_2(t,s_0)&=&\{3,3,6,3,6,3,6,3,6,\cdots\}.
\end{array}
$$
The trajectory on $S_R=\Z_2$ is
$$
\begin{array}{ccl}
x_1(t,\pi(s_0))&=&\{1,2,1,1,2,1,2,1,2,\cdots\},\\
z_2(t,\pi(s_0))&=&\{1,1,2,1,2,1,2,1,2,\cdots\}.
\end{array}
$$

It is ready to verify that
$$
\pi\left(Z_i(t,s_0)\right)=x_i(t,\pi(s_0)),\quad i=1,2.
$$
\end{exa}

\begin{rem}\label{r.5.1.6} Consider the control network (\ref{5.1.11}). If
$$
p_i(0,U(t))=0,\quad i\in [1,n],
$$
that is in each polynomial there is no pure control term, then in Proposition \ref{p.5.1.4} the condition
$u(t)\in S$ can be omitted. Because now we have only $U(t)X(t)$ terms. Since $S$ is an idea,  $U(t)X(t)\in S$ for arbitrary $U(t)$. But if we have pure control term such as $U^k(t)$, it may make the trajectory, starting from $S$, leave $S$ at some time.
\end{rem}

\section{Networks over Product Rings}

\subsection{Product Rings}

\begin{dfn}\label{d.5.4} Let $(R_i,\odot_i,\oplus_i)$, $i=1,2$ be two finite rings.
The product of $R_1$ and $R_2$, denoted by $R=R_1\times R_2$, is defined as follows:
$$
R:=\left\{(r_1,r_2)\;|\;r_1\in R_1,r_2\in R_2\right\}.
$$
The addition and product on $R$ are defined as follows:
\begin{align}\label{5.6}
\begin{array}{ccl}
(r_1,r_2)\odot(s_1,s_2)&:=&(r_1\odot_1 s_1, r_2\odot_2 s_2),\\
(r_1,r_2)\oplus(s_1,s_2)&:=&(r_1\oplus_1 s_1, r_2\oplus_2 s_2),\\
~&~&(r_1,r_2),(s_1,s_2)\in R.
\end{array}
\end{align}
\end{dfn}

According to the definition, it is easy to verify the following result.

\begin{prp}\label{p.5.5} The product set $R=R_1\times R_2$ with operators defined by (\ref{5.6}) is a ring. Moreover,
\begin{align}\label{5.7}
\begin{array}{l}
{\bf 1}_{R}=\left( {\bf 1}_{R_1}, {\bf 1}_{R_2}\right),\\
{\bf 0}_{R}=\left( {\bf 0}_{R_1}, {\bf 0}_{R_2}\right).\\
\end{array}
\end{align}
\end{prp}

Note that if $R=R_1\times R_2$, then
\begin{align}\label{5.8}
|R|=|R_1||R_2|.
\end{align}

Now assume $|R_1|=k_1$, $|R_2|=k_2|$, $\kappa=k_1k_2$. Expressing elements in $R_1$ and $R_2$ into their vector form, we have
$$
\begin{array}{l}
R_1=(1,2,\cdots,k_1-1,0)\sim \d_{k_1}\{1,2,\cdots,k_1\},\\
R_2=(1,2,\cdots,k_1-1,0)\sim \d_{k_1}\{1,2,\cdots,k_1\},\\
\end{array}
$$
Define a mapping $\pi: R_1\times R_2\ra R$ as
$$
\varphi(x_1,x_2):=x_1x_2\in \D_{\kappa}\simeq R.
$$
This is a bijective mapping.

\begin{prp}\label{p.5.6} Assume $R_1=(X,\oplus_1,\odot_1)$ with $|X|=k_1$ and $R_2=(Y,\oplus_2,\odot_2)$ with $|Y|=k_2$. Let $R=R_1\times R_2:=(X\times Y,\oplus, \odot)$. Denote $\varphi(R):=\tilde{R}=\left(W,\tilde{\oplus},\tilde{\odot}\right)$. Then
\begin{align}\label{5.8}
M_{\tilde{\oplus}}=M_{\oplus_1}\left(I_{k_1^2}\otimes M_{\oplus_2}\right)\left(I_{k_1}\otimes W_{[k_2,k_1]}\right).
\end{align}
\begin{align}\label{5.9}
M_{\tilde{\odot}}=M_{\odot_1}\left(I_{k_1^2}\otimes M_{\odot_2}\right)\left(I_{k_1}\otimes W_{[k_2,k_1]}\right).
\end{align}
\begin{align}\label{5.901}
M_{\tilde{\neg}}=M_{\neg_1}\left(I_{k_1}\otimes M_{\neg_2}\right).
\end{align}

\end{prp}

\noindent{\it Proof}.
Consider (\ref{5.8}):
$$
\begin{array}{l}
\varphi[(x_1,y_1)\oplus (x_2,y_2)]=\varphi[(x_1\oplus_1 x_2),(y_1\oplus_2 y_2)]\\
~~~=\varphi[M_{\oplus_1}x_1x_2,M_{\oplus_2}y_1y_2]\\
~~~=M_{\oplus_1}x_1x_2M_{\oplus_2}y_1y_2\\
~~~=M_{\oplus_1}\left(I_{k_1^2}\otimes M_{\oplus_2}\right)x_1x_2y_1y_2\\
~~~=M_{\oplus_1}\left(I_{k_1^2}\otimes M_{\oplus_2}\right)\left(I_{k_1}\otimes W_{[k_2,k_1]}\right)x_1y_1x_2y_2\\
~~~:=M_{\tilde{\oplus}}\varphi((x_1,y_1))\varphi((x_2,y_2)),\\
~~~~~\qquad x_1,x_2\in R_1, y_1,y_2\in R_2.
\end{array}
$$
Since $x_1$, $x_2$, $y_1$, and $y_2$ are arbitrary, (\ref{5.8}) follows.

Similarly, we can prove (\ref{5.9}) and {\ref{5.901}}.

\hfill $\Box$

It is clear that $\varphi:R_1\times R_2 \ra \tilde{R}$ is a ring isomorphism, so hereafter, we do not distinct $R_1\times R_2$ and call $\tilde{R}$ with $\tilde{\oplus}$ and $\tilde{\odot}$ as the product ring and omit the symbol ``$\tilde{~}$" .

 \begin{rem}\label{r.5.601} When elements in $R_i$ are expressed into their vector forms as $x=\d_{k_1}^i\in R_1$ and $y=\d_{k_2}^j\in R_2$, then the element $(x,y)\in R$ is expressed as
\begin{align}\label{5.90101}
(x,y)=\d_{\kappa}^s=\d_{k_1}^i\d_{k_2}^j,
\end{align}
equivalently,
$$
s=(i-1)k_2+j.
$$
This assignment makes formulas  (\ref{5.8})-(\ref{5.901}) available.
\end{rem}

 \begin{exa}\label{e.5.602}
\begin{itemize}
\item[(i)] Consider $\Z^4=\Z_2\times \Z_2$. It is well known that
$$
\begin{array}{l}
M_{+_2}=\d_2[2,1,1,2],\\
M_{\times_2}=\d_2[1,2,2,2].\\
\end{array}
$$
Using formulas (\ref{5.8}) and (\ref{5.9}), we have
\begin{align}\label{5.91101}
\begin{array}{ccl}
M_{\oplus}&=&M_{+_2}\left(I_4\otimes M_{+_2}\right)\left(I_2\otimes W_{[2,2]}\right)\\
~&=&\d_4[4,     3,     2,     1,     3,     4,     1,     2,     2,     1,     4,     3,     1,     2,     3,     4].
\end{array}
\end{align}

\begin{align}\label{5.91102}
\begin{array}{ccl}
M_{\odot}&=&M_{\times_2}\left(I_4\otimes M_{\times_2}\right)\left(I_2\otimes W_{[2,2]}\right)\\
~&=&\d_4[1,     2,     3,     4,     2,     2,     4,     4,     3,     4,     3,     4,     4,     4,     4,     4,
].
\end{array}
\end{align}

Comparing it with the rings in Example \ref{e.3.2} (ii), one sees easily that
$$
\Z_2\times \Z_2=\Z^4 \neq \Z_4.
$$

\item[(ii)] Consider $\Z^6=\Z_2\times \Z_3$. We have
$$
\begin{array}{l}
M_{+_3}=\d_3[2,3,1,3,1,2,1,2,3],\\
M_{\times_3}=\d_3[1,2,3,2,1,3,3,3,3],\\
M_{\neg_3}=\d_3[2,1,3].
\end{array}
$$
Using formulas (\ref{5.8}) and (\ref{5.9}), we have
\begin{align}\label{5.91103}
\begin{array}{ccl}
M_{+^6}&=&M_{+_2}\left(I_4\otimes M_{+_3}\right)\left(I_2\otimes W_{[3,2]}\right)\\
~&=&\d_6[5,6,4,2,3,1,6,4,5,3,1,2,4,5,6,1,2,3,\\
~&~&     2,3,1,5,6,4,3,1,2,6,4,5,1,2,3,4,5,6].
\end{array}
\end{align}

\begin{align}\label{5.91104}
\begin{array}{ccl}
M_{\times^6}&=&M_{\times_2}\left(I_4\otimes M_{\times_3}\right)\left(I_2\otimes W_{[3,2]}\right)\\
~&=&\d_6[1,2,3,4,5,6,2,1,3,5,4,6,3,3,3,6,6,6,\\
~&~&     4,5,6,4,5,6,5,4,6,5,4,6,6,6,6,6,6,6].
\end{array}
\end{align}

\begin{align}\label{5.91105}
\begin{array}{ccl}
M_{\neg^6}&=&M_{\neg_2}\otimes M_{\neg_3}\\
~&=&\d_6[2,     1,     3,     5,     4,     6].
\end{array}
\end{align}
\end{itemize}
\end{exa}

The following proposition is essential for product rings. Using the definition of product ring, it is ready to be verified.

\begin{prp}\label{p.5.7}Let $R=R_1\times R_2$.
Define
\begin{align}\label{5.9a}
\begin{array}{l}
S_1:=\left\{(r_1,{\bf 0}_2)\;|\;r_i\in R_1\right\}\subset R\\
S_2:=\left\{{\bf 0}_1,r_2)\;|\;r_2\in R_2\right\}\subset R\\
\end{array}
\end{align}
Then
\begin{itemize}
\item[(i)] $S_i$, $i=1,2$ are ideals of $R$. Moreover, let
$\pi_i:S_i\ra R_i$ be defined by
\begin{align}\label{5.9b}
\begin{array}{l}
\pi_1\left((r_1,{\bf 0}_2)\right):=r_1,\\
\pi_2\left(({\bf 0}_1,r_2)\right):=r_2.\\
\end{array}
\end{align}
Then $\pi_i$, $i=1,2$ are ring isomorphisms.

\item[(ii)] Let
$\phi_i:R\ra S_i$ be defined by
\begin{align}\label{5.9c}
\begin{array}{l}
\phi_1\left((r_1,r_2)\right):={r_1,\bf 0}_2)\in S_1,\\
\phi_2\left((r_1,r_2)\right):={\bf 0}_1,r_2)\in S_2.\\
\end{array}
\end{align}
Then $\phi_i$ satisfy (\ref{5.1.301}). That is, both $R_1$ and
$R_2$ are proper ideals of $R=R_1\times R_2$.
\end{itemize}
\end{prp}

Recall the corresponding labels for element in product ring and its factor elements, as discussed in Remark \ref{r.5.601}, the following result is obtained by a straightforward computation:

\begin{prp}\label{p.5.701} Let $z=\d_{\kappa}^r\in R=R_1\times R_2$, where $|R_i|=k_i$, $i=1,2$. Then
\begin{itemize}
\item[(i)]
\begin{align}\label{5.90102}
\begin{array}{ccl}
\phi_1(z)&=&\left(I_{k_1}\otimes \J_{k_2}^T\right)z,\\
\phi_2(z)&=&\left(\J_{k_1}^T\otimes I_{k_2}\right)z.\\
\end{array}
\end{align}
\item[(ii)] Numerically, (\ref{5.90102}) can be expressed as
\begin{align}\label{5.90103}
\begin{array}{ccl}
\phi_1(\d_{\kappa}^r)&=&\d_{k_1}^{\a},\\
\phi_2(\d_{\kappa}^r)&=&\d_{k_2}^{\b},\\
\end{array}
\end{align}
where
\begin{align}\label{5.90104}
\begin{array}{ccl}
\a&=&\left[\frac{r-1}{k_2}\right]+1,\\
\b&=&r-(\a-1)k_2.
\end{array}
\end{align}
\end{itemize}
\end{prp}
(We use $[a]$ for the integer part of $a$.)

We also need to emerge elements in $R_1$ or $R_2$ into $R$. Note that
$$
\begin{array}{ccl}
\phi_1^{-1}(\a)&=&(\a,k_2)\in R,\quad \a \in R_1,\\
\phi_2^{-1}(\b)&=&(k_1,\b)\in R,\quad \b \in R_2.\\
\end{array}
$$
Numerically, we have
\begin{align}\label{5.90105}
\begin{array}{ccl}
\phi_1^{-1}(\d_{k_1}^{\a})&=&\d_{\kappa}^p,\\
\phi_2^{-1}(\d_{k_2}^{\b})&=&\d_{\kappa}^q,\\
\end{array}
\end{align}
where
\begin{align}\label{5.90106}
\begin{array}{ccl}
p&=&(\a-1)k_2,\\
q&=&\kappa-k_2+\b.\\
\end{array}
\end{align}

The following proposition shows some properties of product ring.

\begin{prp}\label{p.5.8}~~~
\begin{itemize}
\item[(i)] Let $R_1,~R_2$ be two finite rings. Then
\begin{align}\label{5.902}
R_1\times R_2\cong R_2\times R_1.
\end{align}
\item[(ii)] Assume $R_i\simeq S_i$ $i=1,2$. Then
\begin{align}\label{5.903}
R_1\times R_2\simeq S_1\times S_2.
\end{align}
\item[(iii)] Assume $R_i\cong S_i$ $i=1,2$. Then
\begin{align}\label{5.904}
R_1\times R_2\cong S_1\times S_2.
\end{align}
\end{itemize}
\end{prp}

\noindent{\it Proof}. For (i), let $R^l=R_1\times R_2$ and $R^r=R_2\times R_1$. Define a mapping $\pi:R^l\ra R^r$ as
$$
(r_1,r_2)\mapsto (r_2,r_1),
$$
then it is ready to verify that $\pi$ is an isomorphism. Similar arguments show (ii) and (iii)
\hfill $\Box$

Next, we consider the multiple product of rings
\begin{align}\label{5.905}
R=R_1\times R_2\cdots R_s
\end{align}
can be defined inductively.

It is also ready to verify that
\begin{align}\label{5.906}
R_1\times (R_2\times R_3)=(R_1\times R_2)\times R_3.
\end{align}
Hence the  expression (\ref{5.905}) is legal.

All the properties discussed in this section so far can naturally extended to multi-product rings. To save space, we leave this to the reader.

\begin{dfn}\label{d.5.8} Assume $p_i$, $i=1,2,\cdots,s$ are prime numbers. We define
\begin{align}\label{5.907}
\Z^{\prod_{i=1}^sp_i}:=\Z_{p_1}\times \Z_{p_2}\times \cdots\times \Z_{p_s},
\end{align}
called the prime product ring.
\end{dfn}

For the sake of uniqueness of the expression, in (\ref{5.907}) we need the following assumption:

{\bf Assumption 2}:
$$
p_1\leq p_2\leq \cdots \leq p_s.
$$
Then $\Z^r$, $r\geq 2$ are uniquely defined.

The operators on $\Z^{pq}$ are denoted by $+^{pq}$, $\times^{pq}$, and $\neg^{pq}$ respectively. Note that $\Z^{pq}\neq \Z_{pq}$, say, $\Z^6\neq \Z_6$.

\subsection{Product Networks}

Let $R=R_1\times R_2$ with $|R_i|=k_i$, and $\kappa=k_1k_2$. Assume $m(x)=ax_1^{r_1}x_2^{r_2}\cdots x_{k_1}^{r_{k_1}}$
is a monomial on $R_1$. To merge it into $R$, we have
\begin{align}\label{6.2.1}
\phi_1^{-1}(m(x))=\phi_1^{-1}(a)(k_2)z_1^{r_1}z_2^{r_2}\cdots z_{k_1}^{r_{k_1}}
\end{align}
Note that $k_2=(1,k_2)$, hence
$$
(k_2)z_1^{r_1}z_2^{r_2}\cdots z_{k_1}^{r_{k_1}}=x_1^{r_1}x_2^{r_2}\cdots x_{k_1}^{r_{k_1}}.
$$

Similarly, assume $n(y)=by_1^{r_1}y_2^{r_2}\cdots y_{k_2}^{r_{k_2}}$
is a monomial on $R_2$. To merge it into $R$, we have
\begin{align}\label{6.2.2}
\phi_2^{-1}(n(y))=\phi_2^{-1}(b)(\kappa-k_2+1)z_1^{r_1}z_2^{r_2}\cdots z_{k_1}^{r_{k_1}}
\end{align}

For a polynomial $p$ on $R_i$, we can merge it term by term into $R$. Then we can define the product of networks.

Assume $\Sigma_i\in R_i$, evolving over $R_i$, are defined respectively by

\begin{itemize}
\item[(i)] $\Sigma_1$ over $R_1$ ($|R_1|=k_1$) :
\begin{align}\label{6.2.3}
\begin{array}{ccl}
x_1(t+1)&=&p_1(x_1(t),x_2(t),\cdots,x_m(t)),\\
x_2(t+1)&=&p_2(x_1(t),x_2(t),\cdots,x_m(t)),\\
~&~&\vdots\\
x_m(t+1)&=&p_m(x_1(t),x_2(t),\cdots,x_m(t)).\\
\end{array}
\end{align}
\item[(ii)] $\Sigma_2$ over $R_2$ ($|R_2|=k_2$):
\begin{align}\label{6.2.4}
\begin{array}{ccl}
y_1(t+1)&=&q_1(y_1(t),y_2(t),\cdots,y_n(t)),\\
y_2(t+1)&=&q_2(y_1(t),y_2(t),\cdots,y_n(t)),\\
~&~&\vdots\\
y_n(t+1)&=&q_n(y_1(t),y_2(t),\cdots,y_n(t)).\\
\end{array}
\end{align}
\end{itemize}

Then the product network $\Sigma=\Sigma_1\times \Sigma_2$, evolving over $R=R_1\times R_2$, is defined as follows:

\begin{dfn}\label{d.6.2.1} Let $\Sigma_i$ over $R_i$, $i=1,2$ be defined as above. Define the dynamics of $z_r$ by

\begin{align}\label{6.2.5}
\begin{array}{ccl}
z_r(t+1)&=&\phi_1^{-1}(p_\a(x_1(t),\cdots,x_m(t))\\
~&~&\phi_2^{-1}(q_{\b}(y_1(t),\cdots,y_n(t)), \quad r\in[1,\kappa],
\end{array}
\end{align}
where $\a$ and $\b$ satisfy (\ref{5.90104}).
Then the $\Sigma$, defined by (\ref{6.2.5}) over $R=R_1\times R_2$, is called the product network of $\Sigma_1$ and $\Sigma_2$.
\end{dfn}

We give an example to depict the product network.

\begin{exa}\label{e.6.2.2}
Consider two networks $\Sigma_i$, $i=1,2$ as follows:
\begin{itemize}
\item[(i)] $\Sigma_1$ over $\Z_5$ as:
\begin{align}\label{6.2.6}
\begin{array}{ccl}
x_1(t+1)=3x_1(t)+4x_2(t),\\
x_2(t+1)=2x_1(t)x_2(t).
\end{array}
\end{align}

\item[(ii)] $\Sigma_2$ over $\Z_3$ as:
\begin{align}\label{6.2.7}
\begin{array}{ccl}
y_1(t+1)=2y_1(t)- y_2(t),\\
y_2(t+1)=y_1^2(t).
\end{array}
\end{align}
\end{itemize}

Note that
$$
\begin{array}{l}
\phi_1^{-1}(3)=9,\quad \phi_1^{-1}(4)=12,\\
\phi_1^{-1}(2)=6,\quad  \phi_2^{-1}(2)=14,\\
(1,0) = (1,k_2)=\d_{\kappa}^{k_2}=\d_{15}^3,\\
(0,1)=(k_1,1)=\d_{\kappa}^{\kappa-k_2+1}=\d_{15}^{13}.
\end{array}
$$
On $\Z_3$ we also have
$$
y_1(t+1)=2y_1(t)- y_2(t)=2y_1(t)+2 y_2(t).
$$

Merging $p_1q_1$ into $R$ yields
$$
\begin{array}{l}
\phi_1^{-1}(p_1(t))\phi_2^{-1}(q_1(t))\\
~~~=[9x_1(t)+12x_2(t)][14y_1(t)+14y_2(t)]\\
~~~=9\times 14 x_1(t)y_1(t)+9\times 14 x_1(t)y_2(t)\\
~~~~+12\times 14 x_2(t)y_1(t)+12\times 14 x_2(t)y_2(t)\\
~~~:=6z_1(t)+6z_2(t)+3z_3(t)+3z_4(t).
\end{array}
$$
Merging $p_1q_2$ into $R$ yields
$$
\begin{array}{l}
\phi_1^{-1}(p_1(t))\phi_2^{-1}(q_2(t))
=[9x_1(t)+12x_2(t)][y^2_1(t)]\\
~~~=[9x_1(t)y_1(t)+12x_1(t)y_2(t)]\times 13z_2(t)\\
~~~:=[9z_1(t)+12z_2(t)]\times (13)z_2(t)\\
~~~=[12z_1(t)+6z_2(t)]\times z_2(t)\\
~~~=12z_1(t)z_2(t)+6z^2_2(t).
\end{array}
$$
Merging $p_2q_1$ into $R$ yields
$$
\begin{array}{l}
\phi_1^{-1}(p_1(t))\phi_2^{-1}(q_1(t))=
[6x_1(t)x_2(t)][14y_1(t)+14y_2(t)]\\
~~~=6x_1(t)x_2(t)[y_1(t)+y_2(t)]\\
~~~=6x_1(t)[z_3(t)+z_4(t)]\\
~~~:=18z_1(t) [z_3(t)+z_4(t)]\\
~~~:=3z_1(t)z_3(t)+3z_1(t)z_4(t).\\
\end{array}
$$
Merging $p_2q_2$ into $R$ yields
$$
\begin{array}{l}
\phi_1^{-1}(p_2(t))\phi_2^{-1}(q_2(t))=
[6x_1(t)x_2(t)][y^2_1(t)]\\
~~~=6x_1(t)y_2(t)\times x_2(t)y_2(t)\\
~~~=6z_2(t)z_4(t).\\
\end{array}
$$
Finally, the dynamics of $\Sigma=\Sigma_1\times \Sigma_2$ is described as
\begin{align}\label{6.2.8}
\begin{array}{ccl}
z_1(t+1)&=&6z_1(t)+6z_2(t)+3z_3(t)+3z_4(t)\\
z_2(t+1)&=&12z_1(t)z_2(t)+6z^2_2(t)\\
z_3(t+1)&=&3z_1(t)z_3(t)+3z_1(t)z_4(t)\\
z_4(t+1)&=&6z_2(t)z_4(t).\\
\end{array}
\end{align}
\end{exa}

\begin{rem}\label{r.6.2.3}
\begin{itemize}
\item[(i)] Let $\Sigma_i$ over $R_i$ $i\in [1,s]$ are defined. Then the product network
$\Sigma=\Sigma_1\times \Sigma_2\times\cdots \times \Sigma_s$ can be define in the same way. That is, construct $R=R_1\times R_2\times \cdots \times R_s$ and merge each each $\Sigma_i$ into $\Sigma$. The merging process is exactly the same.
\item[(ii)] More properties of product networks will be discussed later. We will see that the product is a process, which aggregates some networks together.  In next section the decomposition process will be introduced. Finally, we will see that these two processes are mutually invertible.
\end{itemize}
\end{rem}

\section{Decomposition Principle}

Assume $R=R_1\times R_2$, where $|R_i|=k_i$, $i=1,2$, then $|R|=k_1k_2:=\kappa$.
The natural projections $\phi_i:R\ra R_i$, $i=1,2$ have been defined in Definition \ref{d.5.7}. In fact, each $x\in R$ has unique decomposed expression as
$$
x=\d_{\kappa}^r=\d_{k_1}^{\a}\d_{k_2}^{\b},
$$
where $\a$ and $\b$ are as shown in formula (\ref{5.90104}). Moreover,
$$
\phi_1(\d_{\kappa}^r)=\d_{k_1}^{\a},\quad
\phi_2(\d_{\kappa}^r)=\d_{k_2}^{\b}.
$$

\begin{lem}\label{l.6.2} Let $R=R_1\times \R_2$. Then $\phi_i:R\ra R_i$, $i=1,2$  are two ring homomorphisms.
\end{lem}

\noindent{\it Proof}. Let $z_1,z_2\in R =\D_{\kappa}$. In vector form we denote
$$
z_1=\d_{\kappa}^p=\d_{k_1}^{p_1}\d_{k_2}^{p_2},\quad
z_2=\d_{\kappa}^q=\d_{k_1}^{q_1}\d_{k_2}^{q_2}.
$$
Then
$$
z_1\oplus z_2=M_{\oplus}\d_{\kappa}^p\d_{\kappa}^q
=\left[M_{\oplus_1}\d_{k_1}^{p_1}\d_{k_1}^{q_1}\right]\left[M_{\oplus_2}\d_{k_2}^{p_2}\d_{k_2}^{q_2}\right].
$$
Hence
$$
\phi_1(z_1\oplus z_2)=\left[M_{\oplus_1}\d_{k_1}^{p_1}\d_{k_1}^{q_1}\right]
=\phi_1(z_1)\oplus_1 \pi_1(z_2);
$$
and
$$
\phi_2(z_1\oplus z_2)=\left[M_{\oplus_2}\d_{k_2}^{p_2}\d_{k_2}^{q_2}\right]
=\phi_2(z_1)\oplus_1 \pi_2(z_2).
$$
Similarly, we can prove the homomorphic property for product and minus.
\hfill $\Box$

\begin{dfn}\label{d.6.3} Let $R=R_1\times R_2$ be a product finite ring. Let $\Sigma$ be a network over $R$, and $\Sigma_i$, $i=1,2$ be two networks over $R_i$ respectively. $\Sigma$ is said to be the product of $\Sigma_1$ and $\Sigma_2$, denoted by $\Sigma=\Sigma_1\times \Sigma_2$, if for any $z_0\in R$, $z_0=x_0y_0$, where $x_0\in R_1$ and $y_0$ in $R_2$, the trajectories satisfy
\begin{align}\label{6.3}
z(t,z_0)=x(t,x_0)y(t,y_0),\quad t>0.
\end{align}
\end{dfn}

The following theorem, called the decomposition principle, shows that a network over a product ring is decomposable.

\begin{thm}(Decomposition Principle) \label{t.6.4} Assume $\Sigma$ is a (control) network over a product ring $R=R_1\times R_2$.
Define $\Sigma_i:=\phi_i(\Sigma),\quad i=1,2$.
Then
\begin{align}\label{6.4}
\Sigma=\Sigma_1\times \Sigma_2.
\end{align}
\end{thm}

\noindent{\it Proof}. Note that in a network over a finite ring, there are only three operators: addition, product, and minus. Consider an $n$-nod network and let $z_s(t)=x_s(t)y_s(t)$, $s\in [1,n]$. Using Lemma \ref{l.6.2}, it is clear that at each step we have
$$
\begin{array}{l}
z_i(t)\oplus z_j(t)=[x_i(t)\oplus_1 x_j(t)][y_i(t)\oplus_2 y_j(t)],\\
z_i(t)\odot z_j(t)=[x_i(t)\odot_1 x_j(t)][y_i(t)\odot_2 y_j(t)],\\
\neg z_s(t)=[\neg x_s(t)][\neg y_s(t)], \quad i,j,s\in [1,n].\\
\end{array}
$$
Hence, such a decomposition is also available for any polynomial, which contains only these three operators. It follows that
$$
z_s(t+1)=x_s(t+1)y_s(t+1),\quad s\in [1,n].
$$
The conclusion follows.
\hfill $\Box$

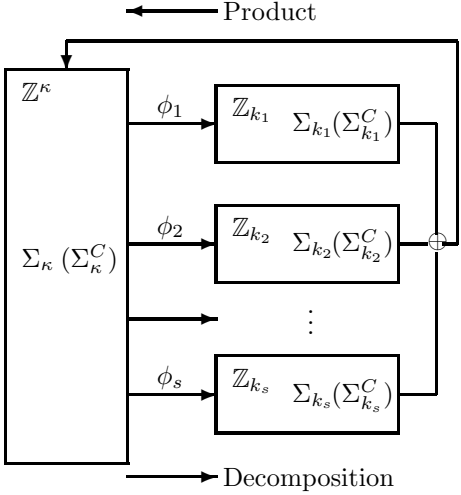
\begin{figure}
\centering
\setlength{\unitlength}{4 mm}
\begin{picture}(17,17)
\thicklines
\put(1,1){\framebox(4,13){}}
\put(8,2){\framebox(6,2.5){}}
\put(8,7){\framebox(6,2.5){}}
\put(8,11){\framebox(6,2.5){}}
\put(5,3.25){\vector(1,0){3}}
\put(5,5.75){\vector(1,0){3}}
\put(5,8.25){\vector(1,0){3}}
\put(5,12.25){\vector(1,0){3}}
\put(6,12.6){$\phi_1$}
\put(6,8.6){$\phi_2$}
\put(6,3.6){$\phi_s$}
\put(1.5,13){$\Z^{\kappa}$}
\put(8.5,12.5){$\Z_{k_1}$}
\put(8.5,8.5){$\Z_{k_2}$}
\put(8.5,3.5){$\Z_{k_s}$}
\put(11,5.25){$\vdots$}
\put(1.5,7.5){$\Sigma_{\kappa}$}
\put(2.8,7.5){($\Sigma^C_{\kappa}$)}
\put(10.5,12){$\Sigma_{k_1}$}
\put(11.8,12){($\Sigma^C_{k_1}$)}
\put(10.5,8){$\Sigma_{k_2}$}
\put(11.8,8){($\Sigma^C_{k_2}$)}
\put(10.5,3){$\Sigma_{k_s}$}
\put(11.8,3){($\Sigma^C_{k_s}$)}
\put(15,8.1){$\oplus$}
\put(14,8.25){\line(1,0){1}}
\put(15.5,8.25){\line(1,0){0.5}}
\put(16,8.25){\line(0,1){6.75}}
\put(16,15){\line(-1,0){13}}
\put(3,15){\vector(0,-1){1}}
\put(14,12.25){\line(1,0){1.3}}
\put(15.3,12.25){\line(0,-1){3.7}}
\put(14,3.25){\line(1,0){1.3}}
\put(15.3,3.25){\line(0,1){4.7}}
\put(5,0.5){\vector(1,0){3}}
\put(8.2,0.2){Decomposition}
\put(8,16){\vector(-1,0){3}}
\put(8.2,15.7){Product}
\end{picture}
\caption{Decomposition of Networks over Rings \label{Fig.8.1}}
\end{figure}

Figure \ref{Fig.8.1} shows the decomposition/product of networks over product rings.

\vskip 5mm

We use an example to demonstrate this theorem.

\begin{exa}\label{e.6.5}
Consider a network over $\Z^4$, defined as follows:
\begin{align}\label{6.5}
\begin{cases}
z_1(t+1)=z_1(t)-z^2_2(t),\\
z_2(t+1)=-z_1^3(t).
\end{cases}
\end{align}
Recall Example \ref{e.5.602} (refer to Example \ref{e.3.2} too). We have
$$
\begin{array}{l}
M_{\oplus}=\d_4[4,3,2,1,3,4,1,2,2,1,4,3,1,2,3,4],\\
M_{\odot}=\d_4[1,2,3,4,2,2,4,4,3,4,3,4,4,4,4,4],\\
M_{\neg}=I_4.
\end{array}
$$
It is easy to calculate that
$$
\begin{array}{l}
z_1(t+1)=M_{\oplus}z_1(t)M_{\neg}M_{\odot} z^2_2(t)\\
~~~=M_{\oplus}z_1(t)M_{\neg}M_{\odot}\PR_4 z_2(t)\\
~~~=M_{\oplus}\left(I_4\otimes (M_{\neg}M_{\odot}PR_4)\right)z(t)\\
~~~:=M_1z(t),
\end{array}
$$
where $M_1$ is calculated as
$$
M_1=\d_4[4,3,2,1,3,4,1,2,2,1,4,3,1,2,3,4].
$$
$$
\begin{array}{l}
z_2(t+1)=M_{\neg}M_{\odot}M_{\odot}z^3_1(t)\\
~~~=M_{\neg}M^2_{\odot}\PR^2_4z_1(t)\\
~~~=M_{\neg}M^2_{\odot}\PR^2_4(I_4\otimes \J_4^T)z(t)\\
~~~:=M_2z(t),
\end{array}
$$
where
$$
M_2=\d_4[1,1,1,1,2,2,2,2,3,3,3,3,4,4,4,4].
$$
Finally, we have the ASSR of (\ref{6.5}) as
\begin{align}\label{6.6}
z(t+1)=Mz(t),
\end{align}
where
\begin{align}\label{6.7}
\begin{array}{ccl}
M&=&M_1*M_2\\
~&=\d_{16}[&13, 9, 5, 1,10,14, 2, 6,\\
~&~        & 7, 3,15,11, 4, 8,12,16].
\end{array}
\end{align}

Next, consider $\Z^4=\Z_2\times \Z_2:=R_1\times R_2$. Splitting $z(t)=x(t)y(t)$ yields

$$
\begin{cases}
x_1(t+1)=L_{11}x(t),\\
x_2(t+1)=L_{12}x(t),\\
\end{cases}
$$
where
$$
L_{11}=\d_2[2,1,1,2],\quad
L_{12}=\d_2[1,1,2,2].
$$
Finally, for projected network over $R_1$ we have
$$
\begin{array}{ccl}
x_1(t+1)&=&L_{11}x_1(t)x_2(t)\\
~&=&L_{11}(I_2\otimes J_2^T\otimes I_2\otimes \J_2^T)x_1(t)y_1(t)x_2(t)y_2(t)\\
~&:=&M_{11}z(t),
\end{array}
$$
where
$$
M_{11}=\d_2[2,2,1,1,2,2,1,1,1,1,2,2,1,1,2,2].
$$

$$
\begin{array}{ccl}
x_2(t+1)&=&L_{12}x_1(t)x_2(t)\\
~&=&L_{12}(I_2\otimes J_2^T\otimes I_2\otimes \J_2^T)x_1(t)y_1(t)x_2(t)y_2(t)\\
~&:=&M_{21}z(t),
\end{array}
$$
where
$$
M_{21}=\d_2[1,1,1,1,1,1,1,1,2,2,2,2,2,2,2,2].
$$

For for projected network over $R_2$ we have
$$
\begin{array}{ccl}
y_1(t+1)&=&L_{11}y_1(t)y_2(t)\\
~&=&L_{11}(\J^T_2\otimes I_2\otimes \J^T_2\otimes I_2)\\
~&~&x_1(t)y_1(t)x_2(t)y_2(t)\\
~&:=&M_{12}z(t),
\end{array}
$$
where
$$
M_{12}=\d_2[2,1,2,1,1,2,1,2,2,1,2,1,1,2,1,2].
$$

$$
\begin{array}{ccl}
y_2(t+1)&=&L_{12}y_1(t)y_2(t)\\
~&=&L_{12}(\J^T_2\otimes I_2\otimes \J^T_2\otimes I_2)x_1(t)y_1(t)x_2(t)y_2(t)\\
~&:=&M_{22}z(t),
\end{array}
$$
where
$$
M_{22}=\d_2[1,1,1,1,2,2,2,2,1,1,1,1,2,2,2,2].
$$

To see that the product of these two trajectories is the trajectory of the original network, we calculate
the product trajectory as
$$
\begin{array}{l}
z^*(t+1)=z_1^*(t+1)z_2^*(t+1)\\
=x_1(t+1)y_1(t+1)x_2(t+1)y_2(t+1)\\
=M_{11}x_1(t)M_{12}y_1(t)M_{21}x_2(t)M_{22}y_2(t)\\
:=M^*z^*(t),\\
\end{array}
$$
where
\begin{align}\label{6.9}
\begin{array}{ccl}
M^*&=&M_{11}*M_{12}*M_{21}*M_{22}\\
 &=\d_{16}[&13, 9, 5, 1,10,14, 2, 6,\\
~&~        &7, 3,15,11, 4, 8,12,16].
\end{array}
\end{align}

Now the $M^*$ obtained by the product of trajectories is exactly the same as the $M$ in (\ref{6.7}), which is obtained by original network.  $M^*=M$ verifies Theorem \ref{t.6.4}.

\end{exa}

\section{Control Networks Over Product Rings}

In a control network the control variables can be considered the same as state variables in polynomials. Hence the decomposition of networks over product rings can be used to control networks directly. Therefore, as a corollary of Theorem \ref{t.6.4}, we have the following result immediately.

\begin{thm}\label{t7.1} Consider a control network $\Sigma^C$ over a product ring $R=R_1\times R_2$. Let $\phi_i:R\ra R_i$ be the natural projections, and $\Sigma^C_i=\phi_i(\Sigma^C)$, $i=1,2$ be the projected networks over $R_i$ respectively.
Then
\begin{itemize}
\item[(i)] $\Sigma^C$ is controllable from $z_0$ to $z_d$, if and only if, $\Sigma^C_i$ is controllable from $\phi_i(z_0)$ to $\phi_i(z_d)$, $i=1,2$.
\item[(ii)] $\Sigma^C$ is control synchronizable to $z_d=\prod_{k=1}^{\kappa}\d_{\kappa}^j$, if and only if,
$\sigma^C_i$ are control synchronizable to $\phi_i(z_d)=\prod_{k=1}^{k_i}\phi_i(\d_{\kappa}^j)$, $i=1,2$.
\item[(iii)] $\Sigma^C$ is observable, if and only if,
$\Sigma^C_i$ are observable, $i=1,2$.
\end{itemize}
\end{thm}

We give an example to demonstrate this.

\begin{exa}\label{e.7.2} Assume $R:=\Z^6=\Z_2\times \Z_3$. A control network over $R$ is
\begin{align}\label{7.1}
\begin{array}{l}
\begin{cases}
z_1(t+1)=z_1^2(t)-4z_2(t),\\
z_2(t+1)=3z_1(t)+u(t),\\
\end{cases}\\
~~~\xi(t)=4z_1(t)+2z_2(t).
\end{array}
\end{align}

The structure matrices of $+^6$, $\times^6$, and $\neg^6$ have been obtained in Example \ref{e.5.602} as (\ref{5.91103}),
(\ref{5.91104}), and (\ref{5.91105}) respectively. Using them, the ASSR of (\ref{7.1}) can be calculated as follows:

$$
\begin{array}{l}
z_1(t+1)=M_{+^6}M_{\times^6}z_1(t)z_1(t)M_{\neg^6}(M_{\times^6}\d_6^4)z_2(t)\\
~~~=M_{+^6}M_{\times^6}\PR_6 z_1(t)M_{\neg^6}(M_{\times^6}\d_6^4)z_2(t)\\
~~~=M_{+^6}M_{\times^6}\PR_6[I_6\otimes (M_{\neg^6}(M_{\times^6}\d_6^4)]z_1(t)z_2(t)\\
~~~=M_{+^6}M_{\times^6}\PR_6[I_6\otimes (M_{\neg^6}(M_{\times^6}\d_6^4)]\\
~~~~~(\J_6^T\otimes I_{36})u(t)z(t)\\
~~~:=L_1u(t)z(t),\\
\end{array}
$$
where
$$
L_1=\d_6[3,3,3,3,\cdots,6,5,4,6]\in {\cal L}_{6\times 216}.
$$

$$
\begin{array}{l}
z_2(t+1)=M_{+^6}u(t)(M_{\times^6}\d_6^3)z_1(t)\\
~~~=M_{+^6}[I_6\otimes(M_{\times^6}\d_6^3)]u(t)z_1(t)\\
~~~=M_{+^6}[I_6\otimes(M_{\times^6}\d_6^3)]\left(I_{36}\otimes \J_6^T\right)u(t)z(t)\\
~~~:=L_2u(t)z(t),\\
\end{array}
$$
where
$$
L_2=\d_6[4,4,4,4,\cdots,6,6,6,6]\in {\cal L}_{6\times 216}.
$$

$$
\begin{array}{l}
y(t)=M_{+^6}(M_{\times^6}\d_6^4)z_1(t)(M_{\times^6}\d_6^2)z_2(t)\\
~~~=M_{+^6}(M_{\times^6}\d_6^4)[I_6\otimes (M_{\times^6}\d_6^2)]z(t)\\
~~~:=Ez(t),\\
\end{array}
$$
where
$$
E=\d_6[3,1,2,6,\cdots,3,4,5,6]\in {\cal L}_{6\times 36}.
$$

Next, we consider the projected networks over $R_1=\Z_2$ and $R_2=\Z_3$ respectively.

\begin{itemize}
\item[(i)] Projected Network over $R_1=\Z_2$:

First, we calculate that
$$
\begin{array}{l}
P_1=(1)=P_1(2)=P_1(3)=1,\\
P_1=(4)=P_1(5)=P_1(0)=0.\\
\end{array}
$$
The projected Network of (\ref{7.1}) over $R_1$ becomes
\begin{align}\label{7.5}
\begin{array}{l}
\begin{cases}
x_1(t+1)=x_1^2(t),\\
x_2(t+1)=x_1(t)+u_1(t),\\
\end{cases}\\
\xi_1(t)=x_2(t).
\end{array}
\end{align}

The ASSR of (\ref{7.5}) is calculated as follows:

$$
\begin{array}{ccl}
x_1(t+1)&=&M_{\times_2}\PR_2\left(\J_2^T\otimes I_2\otimes \J_2^T\right)u_1(t)x(t)\\
~&:=&L_{11}u_1(t)x(t).
\end{array}
$$
where
$$
L_{11}=\d_2[1,1,2,2,1,1,2,2].
$$

$$
\begin{array}{ccl}
x_2(t+1)&=&M_{+_2}\left(I_4\otimes \J_2^T\right)u_1(t)x(t)\\
~&:=&L_{12}u_1(t)x(t),
\end{array}
$$
where
$$
L_{12}=\d_2[2,2,1,1,1,1,2,2].
$$
$$
\begin{array}{ccl}
\xi_1(t)&=&\left(J_2\otimes I_2\right) x(t)\\
~&=&E_1x(t),
\end{array}
$$
where
$$
E_1=\d_2[1,2,1,2].
$$

Finally, we have
$$
x(t+1)=L_1u_1(t)x(t),
$$
where
$$
\begin{array}{ccl}
L_1&=&L_{11}*L_{12}\\
~&=&\d_4[2,2,3,3,1,1,4,4].
\end{array}
$$

\item[(ii)] Projected Network over $R_2=\Z_3$:

First, we have
$$
\begin{array}{l}
P_1(1)=P_2(4)=1,\\
P_2(2)=P_2(5)=2,\\
P_2(3)=P_2(0)=0.\\
 \end{array}
$$
The projected network over $R_2$ is
\begin{align}\label{7.6}
\begin{array}{l}
\begin{cases}
y_1(t+1)=y_1^2(t)-y_2(t),\\
y_2(t+1)=u_2(t),\\
\end{cases}\\
\xi_2(t)=y_1(t)+2y_2(t).
\end{array}
\end{align}

The ASSR of (\ref{7.6}) is calculated as follows:

$$
\begin{array}{ccl}
y_1(t+1)&=&M_{+_3}M_{\times_3}\PR_3[I_3\otimes M_{\neg_3}]\\
~&~&~\left(\J_3^T\otimes I_9\right)u_2(t)y(t)\\
~&:=&L_{21}u_2(t)y(t).
\end{array}
$$
where
$$
L_{21}=\d_3[3,2,1,3,\cdots,1,2,1,3]\in {\cal L}_{3\times 27}.
$$
$$
\begin{array}{ccl}
y_2(t+1)&=&\left(I_3\otimes \J_9^T\right)u_2(t)y(t)\\
~&:=&L_{22}u_2(t)y(t),
\end{array}
$$
where
$$
L_{22}=\d_3[1,1,1,1,\cdots,3,3,3,3]\in {\cal L}_{3\times 27}.
$$

$$
\begin{array}{ccl}
\xi_2(t)&=&M_{+_3}\left(I_3\otimes (M_{\times_3}\d_3^2)\right) y(t)\\
~&:=&E_2y(t),
\end{array}
$$
where
$$
E_2=\d_3[3,2,1,1,3,2,2,1,3].
$$

Finally, we have
$$
y(t+1)=L_2u_2(t)y(t),
$$
where
$$
\begin{array}{ccl}
L_2&=&L_{21}*L_{22}\\
~&=&\d_9[7,4,1,7,\cdots,3,6,3,9]\in {\cal L}_{9\times 27}.
\end{array}
$$
\end{itemize}

Finally, we consider the observability of network (\ref{7.1}). According to Theorem \ref{t7.1}, it is enough to verify the observability of sub-networks (\ref{7.5}) and (\ref{7.6}).

\begin{itemize}

\item[(i)] Consider sub-network (\ref{7.5}) again. We use the technique of set controllability of auxiliary network proposed by \cite{che18}, for this.

Following \cite{che18}, we construct the auxiliary network as
\begin{align}\label{7.7}
\begin{cases}
x(t+1)=L_1u(t)x(t)\\
x^*(t+1)=L_1u(t)x^*(t).
\end{cases}
\end{align}
Set $w(t)=x(t)x^*(t)$, then the ASSR of (\ref{7.7}) becomes
\begin{align}\label{7.8}
\begin{array}{ccl}
w(t+1)&=&L_1u(t)x(t)L_1u(t)x^*(t)\\
~&=&L_1\left(I_{2^3}\otimes L_1\right)u(t)x(t)u(t)x^*(t)\\
~&=&L_1\left(I_{2^3}\otimes L_1\right)\left(I_3\otimes W_{[2,4]}\right)u(t)^2x(t)x^*(t)\\
~&=&L_1\left(I_{2^3}\otimes L_1\right)\left(I_3\otimes W_{[2,4]}\right)PR_2u(t)w(t)\\
~&:=&\Psi u(t)w(t),
\end{array}
\end{align}
where
$$
\Phi=L_1\left(I_{2^3}\otimes L_1\right)\left(I_3\otimes W_{[2,4]}\right)PR_2.
$$

Using $E_1$, the output distinguishable pairs are
$$
\begin{array}{llll}
\{(\d_4^1,\d_4^2),& (\d_4^1,\d_4^4),& (\d_4^2,\d_4^1),&(\d_4^2,\d_4^3),\\
(\d_4^3,\d_4^2),&(\d_4^3,\d_4^4),&(\d_4^4,\d_4^1),&(\d_4^4,\d_4^3).\}.
\end{array}
$$
Their corresponding $W(t)=x(t)x^*(t)$ are
$$
W=\d_{16}\{2,4,5,7,10,12,13,1,5\}.
$$

The set controllability matrix of $\D_{16}$ to $W$ can be calculated as follows:
First, set
$$
M=L1\d_2^1+_{{\cal B}}L1\d_2^2.
$$
Then the controllability matrix of (\ref{7.8}) is calculated as
$$
{\cal C}^1={\dsum_{{\cal B}}}_{i=1}^{16}M^{(i)}.
$$
The set controllability matrix of (\ref{7.8}) to $W$ is
$$
\begin{array}{l}
{\cal C}^1_W=ID_{W}{\cal C}\\
~~~=[0,0,1,1,0,0,1,1,1,1,0,0,1,1,0,0].
\end{array}
$$

So the sub-network (\ref{7.5}) is not observable.
Precisely speaking, the following pairs are indistinguishable.
$$
S_{id}^1=\left\{(\d_4^1,\d_4^2),(\d_4^3,\d_4^4)\right\}.
$$

\item[(ii)] Consider sub-network (\ref{7.6}). A similar argument shows that
$$
\begin{array}{ccl}
{\cal C}^2_W&=[&0,1,1,0,1,1,1,1,0,\\
~&~            &1,0,1,1,0,1,0,1,1,\\
~&~            &1,1,0,1,1,0,1,0,1,\\
~&~            &0,1,1,0,1,1,1,1,0,\\
~&~            &1,0,1,1,0,1,0,1,1,\\
~&~            &1,1,0,1,1,0,1,0,1,\\
~&~            &1,0,1,1,0,1,0,1,1,\\
~&~            &1,1,0,1,1,0,1,0,1,\\
~&~            &0,1,1,0,1,1,1,1,0].
\end{array}
$$
It follows that the sub-network (\ref{7.6}) is not observable. Moreover, the indistinguishable
pairs are
$$
\begin{array}{ccl}
S_{id}^2&=\{&(\d_9^1,\d_9^4), (\d_9^1,\d_9^9),(\d_9^2,\d_9^5),\\
~&~&         (\d_9^2,\d_9^7),(\d_9^3,\d_9^6), (\d_9^3,\d_9^8), \\
~&~&         (\d_9^4,\d_9^9),(\d_9^5,\d_9^7), (\d_9^6,\d_9^8)\}. \\
\end{array}
$$

\end{itemize}

According to Theorem \ref{t7.1}, the network (\ref{7.1}) is not observable.
Moreover,
$\left(\d_{36}^i,\d_{36}^j\right)$ is an indistinguishable pair, if and only if,
$$
\d_{36}^{i}=\d_2^{\a(i)}\d_3^{\b(i)},\quad \d_{36}^{j}=\d_2^{\a(j)}\d_3^{\b(j)},
$$
where
$$
\left(\d_2^{\a(i)},\d_2^{\a(j)}\right)\in S_{id}^1,
$$
or
$$
\left(\d_3^{\b(i)},\d_3^{\b(j)}\right)\in S_{id}^2.
$$

\end{exa}

\section{Linear Networks}

In this section we consider  linear (control) networks, which were the main object concerned by most existing literature, where the controls were ignored. \cite{li16,li19,men20}.

\begin{dfn}\label{d.10.1} Let $A\in \Z^{\kappa}_{n\times n}$ and $\Z^{\kappa}=\prod_{i=1}^s\Z_{k_i}$. Denote $A=(a_{i,j})$ where $a_{i,j}\in \Z^{\kappa}$. Then the projection $\phi_i: \Z^{\kappa}_{n\times n}\ra \Z_{k_i}^{n\times n}$ is defined by
\begin{align}\label{10.1}
\phi_i(A):=A^i\in \Z_{k_i}^{n\times n},\quad i=1,2,\cdots,s,
\end{align}
where
$A^i=(\phi_i(a_{i,j}))$.
\end{dfn}

Consider a linear control network over $\Z^{\kappa}$ as
\begin{align}\label{10.2}
\begin{cases}
Z(t+1)=AZ(t)+^{\kappa} BU(t),\\
Y(t)=CZ(t),
\end{cases}
\end{align}
where
$$
\begin{array}{l}
Z(t)=(Z_1(t),Z_2(t),\cdots,Z_n(t))^T\in \Z^{\kappa}_n,\\
U(t)=(U_1(t),U_2(t),\cdots,U_m(t))^T \in \Z^{\kappa}_m,\\
Y(t)=(Y_1(t),Y_2(t),\cdots,Y_p(t))^T \in \Z^{\kappa}_p,\\
A\in \Z^{\kappa}_{n\times n},\quad
B\in \Z^{\kappa}_{n\times m},\quad
C\in \Z^{\kappa}_{p\times n}.
\end{array}
$$

Using decomposition principle, the following result is obvious.

\begin{prp}\label{p.10.1} Assume $\kappa=\prod_{i=1}^sk_i$, where $k_i$, $i=1,2,\cdots,s$ are prime numbers. Then there exist sub-linear networks
\begin{align}\label{10.3}
\begin{cases}
X^i(t+1)=A^iX^i(t)+_{k_i} B^iU^i(t),\\
Y^i(t)=C^iX^i(t),\quad, i=1,2,\cdots,s,
\end{cases}
\end{align}
where
$$
\begin{array}{l}
A^i=\phi_i(A)\in {\Z_{k_i}}_{n\times n},\\
B^i=\phi_i(B)\in {\Z_{k_i}}_{n\times m},\\
C^i=\phi_i(C)\in {\Z_{k_i}}_{p\times n},
\end{array}
$$
such that
\begin{align}\label{10.4}
\Sigma=\Sigma_1\times \Sigma_2\times \cdots\times \Sigma_s.
\end{align}

Then the system $\Sigma$ is controllable, or observable, if and only if, every factor sub-systems are controllable, or observable respectively.
\end{prp}

According to Proposition \ref{p.10.1}, various control problems of $\Sigma$ can be solved through $\Sigma_i$, $i\in[1,s]$.

Next, we consider how to solve control problems for each $\Sigma_i$.

Consider network (\ref{10.3}) . Now assume it is defined over $\Z_k$, where $k=k_i$ is a prime number.

Then the ASSR of $X_i$ can be obtained as follows:

For free terms:
$$
\begin{array}{ccl}
\Row_i(A^i) x(t)&=&M^{n-1}_{+_k} M_{\times_k}\d_k^{a_{i,1}}x_1(t) M_{\times_k}\d_k^{a_{i,2}}x_2(t)\\
 ~&~&\cdots M_{\times_k}\d_k^{a_{i,n}}x_n(t)\\
~&=&M_{\times_k}M^{n-1}_{+_k}\d_k^{a_{i,1}}\left(I_k\otimes \d_k^{a_{i,2}}\right)\\
~&~&\left(I_{k^2}\otimes \d_k^{a_{i,3}}\right)\cdots \left(I_{k^{n-1}}\otimes \d_k^{a_{i,n}}\right)x(t)\\
~&:=&A_ix(t),\quad i=1,2,\cdots,n.
\end{array}
$$
For control terms:
$$
\begin{array}{ccl}
\Row_i(B^i)u(t)&=&M^{m-1}_{+_p} M_{\times_k}\d_k^{b_{i,1}}u_1(t) M_{\times_k}\d_k^{b_{i,2}}u_2(t)\\
 ~&~&\cdots M_{\times_k}\d_k^{b_{i,m}}u_m(t)\\
~&=&M_{\times_k}M^{m-1}_{+_k}\d_k^{b_{i,1}}\left(I_k\otimes \d_k^{b_{i,2}}\right)\\
~&~& \left(I_{k^2}\otimes \d_k^{b_{i,3}}\right)\cdots \left(I_{k^{m-1}}\otimes \d_k^{b_{i,m}}\right)u(t)\\
~&:=&B_iu(t),\quad i=1,2,\cdots,n.
\end{array}
$$
Finally, we have component-wise ASSR as
\begin{align}\label{10.5}
\begin{array}{ccl}
x_i(t+1)&=&M_{+_k} B_iu(t)A_ix(t)\\
~&=&M_{+_k}B_i\left(I_{k^m}\otimes A_i\right) u(t)x(t)\\
~&:=&L_iu(t)x(t),\quad i=1,2,\cdots,n.
\end{array}
\end{align}
The overall ASSR is
\begin{align}\label{10.6}
x(t+1)=Lu(t)x(t),
\end{align}
where
$$
L=L_1*L_2*\cdots *L_n.
$$
Similarly, for outputs we have
$$
\begin{array}{ccl}
y_j(t)&=&M^{n-1}_{+_k} M_{\times_k}\d_k^{c_{j,1}}x_1(t) M_{\times_k}\d_k^{c_{j,2}}x_2(t) \\
~&~&\cdots M_{\times_k}\d_k^{a_{j,n}}x_n(t)\\
~&=&M_{\times_k}M^{n-1}_{+_k}\d_k^{c_{j,1}}\left(I_k\otimes \d_k^{c_{j,2}}\right)\left(I_{k^2}\otimes \d_k^{c_{j,3}}\right)\\
~&~&\cdots \left(I_{k^{n-1}}\otimes \d_k^{a_{i,n}}\right)x(t)\\
~&:=&E_jx(t),\quad j=1,2,\cdots,p.
\end{array}
$$
And
\begin{align}\label{10.7}
y(t)=Ex(t),
\end{align}
where
$$
E=E_1*E_2*\cdots*E_p.
$$

We consider an example.
\begin{exa}\label{e.10.2}
Consider a linear network $\Sigma$ over $\Z^6$ as
\begin{align}\label{10.8}
\begin{array}{cl}
\Sigma:&Z(t+1)=AZ(t)+^6 BU(t),\\
~&\Xi(t)=CZ(t),
\end{array}
\end{align}
where
$$
A=\begin{bmatrix}
3&4\\
1&5
\end{bmatrix},\quad
B=\begin{bmatrix}
3\\
2
\end{bmatrix},\quad
C=\begin{bmatrix}
2&3
\end{bmatrix}.
$$

Decomposing (\ref{10.8}) into $\Z_2$ and $\Z_3$ yields
\begin{itemize}
\item[(i)] $\Sigma_1$ over $\Z_2$ as
\begin{align}\label{10.9}
\begin{array}{cl}
\Sigma_1:&X(t+1)=A^1X(t)+B^1V(t),\\
~&\Xi^1(t)=C^1X(t),
\end{array}
\end{align}
where
$$
A^1=\begin{bmatrix}
1&1\\
1&0
\end{bmatrix},\quad
B^1=\begin{bmatrix}
1\\
0
\end{bmatrix},\quad
C^1=\begin{bmatrix}
0&1
\end{bmatrix}.
$$

\item[(ii)] $\Sigma_2$ over $\Z_3$ as
\begin{align}\label{10.10}
\begin{array}{cl}
\Sigma_2:&Y(t+1)=A^2Y(t)+B^2W(t),\\
~&\Xi^2(t)=C^2Y(t),
\end{array}
\end{align}
where
$$
A^2=\begin{bmatrix}
0&1\\
1&2
\end{bmatrix},\quad
B^2=\begin{bmatrix}
0\\
2
\end{bmatrix},\quad
C^2=\begin{bmatrix}
2&0
\end{bmatrix}.
$$
\end{itemize}

Then the control problems of $\Sigma$ can be solved via
$\Sigma_1$ and $\Sigma_2$.

In the following we consider the controllability only.

\begin{itemize}
\item[(i)] For $\Sigma_1$:
It is easy to calculate that
$$
\begin{array}{ccl}
L^1_1&=&\d_2[2,2,1,1,1,1,2,2],\\
L^1_2&=&\d_2[2,1,1,2,2,1,1,2],\\
L^1&=&L^1_1*L^1_2\\
~&=&\d_4[4,3,1,2,2,1,3,4],\\
\end{array}
$$
Then we have
$$
\begin{array}{ccl}
M_1&:=&L^1\d_2^1+L^1\d_2^2\\
~&=&\begin{bmatrix}
0&     1&     1&     0\\
1&     0&     0&     1\\
0&     1&     1&     0\\
1&     0&     0&     1
\end{bmatrix}
\end{array}
$$
The controllability matrix is
$$
\begin{array}{ccl}
{\cal C}^1&=&{\dsum_{{\cal B}}}_{i=1}^4 M_1^{(i)}\\
~&=&\J_{4\times 4}.
\end{array}
$$
Hence, $\Sigma_1$ is completely controllable.

\item[(ii)] For $\Sigma_2$:
We calculate that
$$
\begin{array}{ccl}
L^2_1&=&\d_3[1,2,3,1,2,3,1,2,3,1,2,3,1,\\
~&~&         2,3,1,2,3,1,2,3,1,2,3,1,2,3].\\
L^2_2&=&\d_3[2,1,3,3,2,1,1,3,2,1,3,2,2,\\
~&~&         1,3,3,2,1,3,2,1,1,3,2,2,1,3].\\
L^2&=&L^2_1*L^2_2\\
~&=&    \d_9[2,4,9,3,5,7,1,6,8,1,6,8,2,\\
~&~&         4,9,3,5,7,3,5,7,1,6,8,2,4,9].
\end{array}
$$
Then we have
$$
\begin{array}{ccl}
M_2&:=&L^2\d_3^1+L^2\d_3^2+L^2\d_3^3\\
~&=&\begin{bmatrix}
1&0&0&1&0&0&1&0&0\\
1&0&0&1&0&0&1&0&0\\
1&0&0&1&0&0&1&0&0\\
0&1&0&0&1&0&0&1&0\\
0&1&0&0&1&0&0&1&0\\
0&1&0&0&1&0&0&1&0\\
0&0&1&0&0&1&0&0&1\\
0&0&1&0&0&1&0&0&1\\
0&0&1&0&0&1&0&0&1\\
\end{bmatrix}
\end{array}
$$
The controllability matrix is
$$
{\cal C}^2={\dsum_{{\cal B}}}_{i=1}^9 M_2^{(i)}=\J_{9\times 9}.
$$
Hence, $\Sigma_2$ is completely controllable.

\end{itemize}

According to Proposition \ref{p.10.1}, we conclude that the original network $\Sigma$ is completely controllable.
\end{exa}

\section{Representation Theorem}

The purpose of this section is to show that all the $\kappa$-valued logical networks can be expressed as networks over finite ring  $\Z^{\kappa}$.
We present this result as the representation theorem. The importance of this result lies on that the technique developed in this paper is universally applicable for any networks over finite set.

To begin with, we consider Boolean (or two-valued) logic. Note that
$$
X\wedge Y=X\times_2 Y,\quad X,Y\in {\cal D}_2,
$$
and
$$
\neg X=X+_2 1,\quad X\in {\cal D}_2.
$$
Hence, both $\wedge$ and $\neg$ can be expressed by operators in $\Z_2$. Since $(\wedge,\neg)$ is an adequate set (generators) of Boolean logic \cite{ham88}, any Boolean function can be expressed by $\{+_2,\times_2\}$.

This factor is also true for $k$-valued logic. The following theorem, called the representation theorem, shows that
all $\kappa$-values networks can be expressed as networks over $\Z^{\kappa}$.

First, we consider a special case when the cardinal number of a finite set $S$ is  prime.

\begin{lem}\label{l.8.1} Consider a finite set  $S$ with $|S|=k$ is a prime number. then a network over $S$ can be expressed as a network over $\Z_{k}$.
\end{lem}

\noindent {\it Proof}.
A minimum adequate set of $k$ valued logic has been constructed through \cite{che20,chepr,chepr2} as
$\{\varphi,\gamma\}$, where $\varphi$ is a binary operator and $\gamma$ is a unary operator with their structure matrices $M_{\varphi}$ and $M_{\gamma}$ respectively as follows:

\begin{align}\label{8.1}
M_{\varphi}=[M_1,M_2,\cdots,M_{k}],
\end{align}
where
$$
\begin{array}{l}
M_1=\d_{k}[1,2,3,\cdots,k]M_{\sigma},\\
M_2=\d_{k}[2,2,3,\cdots,k]M_{\sigma},\\
\cdots\\
M_k=\d_{k}[k,k,k,\cdots,k]M_{\sigma},\\
\end{array}
$$
and $\sigma=(1,2,\cdots,k)\in {\bf S}_{k}$, (${\bf S}_{k}$ is the symmetric group over $k$ objects\cite{hun74}).
$$
M_{\gamma}=\d_{k}[1,1,2,3,\cdots,k-1].
$$
Hence, it is enough to
prove that both $\varphi$ and $\gamma$ can be expressed by two polynomials over $\Z_{k}$ respectively.

Define a set of index functions
\begin{align}\label{8.2}
\Gamma_{\a}(x):=\prod_{j\neq \a}(\a-j)^{-1}(x-j),\quad \a\in [0,k-1],
\end{align}
Note that since $\Z_k$ is a field, $(\a-j)^{-1}$ for $j\neq \a$ is properly defined.
Then it is ready to verify that
$$
\Gamma_{\a}(x)=\begin{cases}
1,\quad x=\a,\\
0,\quad x\neq \a.
\end{cases}
$$

Define
$$
\begin{array}{ccl}
P_{\gamma}(x)&:=&\Gamma_1(x)\times_{k}1\\
~&+_{k}&\Gamma_2(x)\times_{k}1\\
~&+_{k}&\Gamma_3(x)\times_{k}2\\
~&+&\cdots\\
~&+&\Gamma_{k}(x)\times_{k}(k-1).\\
\end{array}
$$
Then it is clear that
$$
\gamma(x)=P_{\gamma}(x).
$$

Similarly, define
$$
\begin{array}{ccl}
P_{\varphi}(x,y)&:=&{+_{\kappa}}_{i=1}^{\kappa} {+_{\kappa}}_{j=1}^{\kappa}
 \Gamma_i(x)\Gamma_j(y)\varphi(i,j)\\
~&=&\varphi(x,y).
\end{array}
$$
\hfill $\Box$

Next, we consider general case. Assume $\kappa =k_1k_2\cdots k_s$ is the prime number decomposition of $\kappa$.
$\Sigma$ is a network over $S$ with $|S|=\kappa$. Denote by $S=\{1,2,\cdots,\kappa-1,0\}$. Using vector form expression
$$
\vec{i}=\begin{cases}
\d_{\kappa}^i,\quad i\neq 0,\\
\d_{\kappa}^{\kappa},\quad i=0.
\end{cases}
$$
Set
$$
\D_{\kappa}:=\left\{\d_{\kappa}^1,\d_{\kappa}^2,\cdots,\d_{\kappa}^{\kappa}\right\}.
$$
Then the dynamics of $\Sigma$ becomes a mapping $\Sigma:\D_{\kappa}\ra \D_{\kappa}$. As a $\kappa$-valued logical network, its dynamics can be expressed as
\begin{align}\label{8.3}
x(t+1)=Lx(t),\quad x(t)\in \D_{\kappa},
\end{align}
where $L\in {\cal L}_{\kappa\times \kappa}$.

Set
$$
\kappa_i:=
\begin{cases}
1,\quad i=1,\\
\prod_{j=1}^{i-1}k_j,\quad 2\leq i\leq s.\\
\end{cases}
$$
$$
\kappa^i:=
\begin{cases}
1,\quad i=s,\\
\prod_{j=i+1}^{s}k_j,\quad 1\leq i\leq s-1.\\
\end{cases}
$$

Construct a set of mappings $E_i:\D_{\kappa}\ra \D_{k_i}$ as
\begin{align}\label{8.4}
E_i=\J^T_{\kappa_i}\otimes I_{k_i} \otimes \J^T_{\kappa^i},\quad i\in[1,s].
\end{align}


%
%

Now we are ready to present the representation theorem:

\begin{thm}\label{t.8.2} (Representation Theorem) A $\kappa$-valued network can be expressed as a network over $\Z^{\kappa}$.
\end{thm}

\noindent {\it Proof}.

Let $x=(x_1,x_2,\cdots,x_s)\in \Z^{\kappa}$, where $x_i\in \Z_{k_i}$. Recall the projections $\phi_i: x\mapsto x_i$, $i\in [1,s]$. It is obvious that $E_i$ defined in (\ref{8.4}) is the structure matrix of $\phi_i$.

Define
\begin{align}\label{8.6}
\Psi^i_{\ell}(x):=\Gamma_{\ell}(\phi_i(x))\J_{\kappa}\in \D_{\kappa},
\quad \ell\in \D_{k_i},\; i\in [1,s],
\end{align}
where $\Gamma_{\ell}: \D_{k_i}\ra \{0,1\}$ is defined by (\ref{8.2}).
It follows that

\begin{align}\label{8.7}
\Psi^i(x)=\begin{cases}
\J_{\kappa},\quad x_i=\ell,\\
{\bf 0}_{\kappa},\quad x_i\neq \ell.
\end{cases}
\end{align}

Finally, (\ref{8.3}) can be expressed as

\begin{align}\label{8.8}
\begin{array}{ccl}
x(t+1)&=&\left[{+^{\kappa}}_{\ell_1\in \D_{k_1}}{+_{\kappa}}_{\ell_2\in \D_{k_2}}\cdots {+_{\kappa}}_{\ell_s\in \D_{k_s}}\right.\\
~&~&\left.{\times_{\kappa}}_{i=1}^s\Psi^i_{\ell_i}(x)M\ell_1 \ell_2 \cdots \ell_s\right]x(t).
\end{array}
\end{align}

\hfill $\Box$

\begin{rem}\label{r.8.3}
\begin{itemize}

\item[(i)] Since the proof of Theorem \ref{t.8.2} is constructive, it can also be used to convert a general finite network into a network over $\Z^{\kappa}$.

\item[(ii)] Since a polynomial over $\Z^{\kappa}$ can be decomposed into a set of polynomials over $\Z_{p_i}$, $p_i|\kappa$. Consider any network $\Sigma$ over $S$ over $S$ for $S={\cal D}_{\kappa}$. If $\kappa =p_1P_2,\cdots, p_s$, then the network  $\Sigma$ can be decomposed into $\Sigma_i$, $i=1,2,\cdots,s$, where $\Sigma_i$ is a sub-network over $\Z_{p_i}$. Then the results obtained for networks over finite fields can be used for each sub-networks.
\end{itemize}
\end{rem}

\begin{exa}\label{e.8.4} Consider a network over $R$ with $|R|=6$. Its dynamics  is described by
\begin{align}\label{8.9}
z(t+1)=Mz(t),
\end{align}
where
$$
M=\d_6[4 6 1 3 2 5].
$$

Split $z=xy$, where $x\in \Z_2$, $y\in \Z_3$.

To identify $x=1$ we construct
$$
\Gamma_1(\phi_1(x)):=(1-0)^{-1}(x-0)=x=\phi_1(z).
$$
Similarly, to identify $x=0$ we construct
$$
\Gamma_0(\phi_1(x)):=(0-1)^{-1}(x-1)=x-1=(\phi_1(z)-1).
$$
To identify $y=1$ we construct
$$
\Gamma_1(\phi_2(x)):=2(1-0)^{-1}(y-0)(1-2)^{-1}(y-2)\phi_2(z)(\phi_2(z)-2).
$$
To identify $y=2$ we construct
$$
\Gamma_2(\phi_2(x)):=2(2-0)^{-1}(y-0)(2-1)^{-1}(y-1)\phi_2(z)(\phi_2(z)-1).
$$
To identify $y=0$ we construct
$$
\Gamma_0(\phi_2(x)):=2(0-1)^{-1}(y-1)(0-2)^{-1}(y-2)(\phi_2(z)-1)(\phi_2(z)-2).
$$
Finally, we have
\begin{align}\label{8.10}
\begin{array}{ccl}
z(t+1)&=&\left[\left(\phi_1(z){\bf 1}_6\right)\times^6\right.\\
~&~&\left.\left(2\phi_2(z)(\phi_2(z)-2){\bf 1}_6\right)\times^6 4\right]\\
~&+^6&\left[\left(\phi_1(z){\bf 1}_6\right)\times^6\left(2(2-0)^{-1}(y-0)\right.\right.     \\
~&~&\left.\left.(2-1)^{-1}(y-1)\phi_2(z)(\phi_2(z)-1){\bf 1}_6\right)\times^6 6\right]\\
~&+^6&\left[\left(\phi_1(z){\bf 1}_6\right)\times^6\left(2(2-0)^{-1}(y-0)\right.\right.     \\
~&~&\left.\left.(2-1)^{-1}(y-1)\phi_2(z)(\phi_2(z)-1){\bf 1}_6\right)\times^6 1\right]\\
~&=&\left[\left((\phi_1(z)-1){\bf 1}_6\right)\times^6\right.\\
~&~&\left.\left(2\phi_2(z)(\phi_2(z)-2){\bf 1}_6\right)\times^6 4\right]\\
~&+^6&\left[\left((\phi_1(z)-1){\bf 1}_6\right)\times^6\left(2(2-0)^{-1}(y-0)\right.\right.     \\
~&~&\left.\left.(2-1)^{-1}(y-1)\phi_2(z)(\phi_2(z)-1){\bf 1}_6\right)\times^6 6\right]\\
~&+^6&\left[\left((\phi_1(z)-1){\bf 1}_6\right)\times^6\left(2(2-0)^{-1}(y-0)\right.\right.     \\
~&~&\left.\left.(2-1)^{-1}(y-1)\phi_2(z)(\phi_2(z)-1){\bf 1}_6\right)\times^6 1\right]\\
\end{array}
\end{align}
\end{exa}

\begin{rem}\label{r.8.5} From the proof of Theorem \ref{t.8.2} (or Example \ref{e.8.4}) one sees easily that in the representation of arbitrary finite network to networks over $\Z^{\kappa}$ the operators over factor rings $\Z_{k_i}$
are also used. Therefore, precisely speaking, an arbitrary finite network (with $\kappa$ elements) can be expressed into a network over $\Z^{\kappa}$ and its factor rings. This is because the projections $\phi_i:\Z^{\kappa}\ra \Z_{k_i}$ are used in the presentation. But it does not matter in most applications. For instance, decomposition principle is still available. Because in the decomposing process, the projection becomes either identity, if the projected sub-ring coincides with the object ring of the decomposition, or zero,  if the projected sub-ring is not the object ring of the decomposition.
\end{rem}

\section{Conclusion}

In this paper the networks over finite rings are proposed and investigated. Using STP, the algebraic conditions are obtained for verifying whether a finite set with two operators is a (commutative) ring, which provides a method to construct more finite (commutative) rings. Then the networks over finite rings are investigated. The networks over finite rings are obvious generalization of the networks over finite fields discussed in some current literature, and it significantly enlarged the application area of finite networks. It is also shown that all existing STP based technique developed for multi and mix valued logical (control) networks are applicable to networks over finite rings.

Then we show that if the bearing finite ring of a network has a proper idea, then  a sub-network can be obtained. This sub-network is exactly an invariant subspace for the dynamics of the original (control) network.

Moreover, the product rings are introduced. Its properties are explored. A set of important product rings, denoted by $\Z^{\kappa}$ are introduced, which is the product of $\Z_{k_i}$, where $k_i$ are prime facts of $\kappa$. Using  product rings, the decomposition of networks over product rings into its factor rings is revealed. This decomposition theorem is called the decomposition principle. Decomposition principle can reduce the computational complexity of networks over finite rings by investigating sub-networks over factor rings to analyse and control the overall network. Then the control problems for networks over finite rings are investigated via its factor sub-networks using decomposition principle. Particularly, the linear control networks over finite rings are analyzed in detail.

Finally, the representation theorem is presented, which shows that any finite network over finite set $S$ with $|S|=\kappa$ can be expressed as a network over product ring $\Z^{\kappa}$. This result is powerful, because it claims that the technique developed in this paper for networks over finite rings is applicable to arbitrary finite networks.

For the sake of applications, the most challenging and urgent issue to be solved is to develop an easily computable technique to convert an (arbitrary) finite network into a network over $\Z^{\kappa}$.

\end{document}